\documentclass[11pt,preprint]{aastex}
\pdfoutput=1
\usepackage{natbib}
\usepackage{color}
\usepackage{amssymb}
\usepackage{amsmath}
\usepackage{graphicx}

\shorttitle{Transit Duration Distribution from \Kepler}
\shortauthors{Moorhead et al. }

\begin{document}

\title{The Distribution of Transit Durations for {\em Kepler} Planet Candidates and Implications for their Orbital Eccentricities}

\author{
Althea V. Moorhead\altaffilmark{1},
Eric B. Ford\altaffilmark{1}, 
Robert C. Morehead\altaffilmark{1}, 
Jason Rowe\altaffilmark{2,3}, 
William J. Borucki\altaffilmark{3}, 
Natalie M. Batalha\altaffilmark{6}, 
Stephen T. Bryson \altaffilmark{3},
Douglas A. Caldwell\altaffilmark{3,7}, 
Daniel C. Fabrycky\altaffilmark{4}, 
Thomas N. Gautier III\altaffilmark{8},
David G. Koch\altaffilmark{3}
Matthew J. Holman\altaffilmark{5}, 
Jon M. Jenkins\altaffilmark{2,3}
Jie Li\altaffilmark{2,3}
Jack J. Lissauer\altaffilmark{3}, 
Philip Lucas\altaffilmark{9},
Geoffrey W. Marcy\altaffilmark{13},
Samuel N. Quinn\altaffilmark{5},
Elisa Quintana \altaffilmark{2,3}
Darin Ragozzine\altaffilmark{5}, 
Avi Shporer\altaffilmark{13,12}, 
Martin Still \altaffilmark{3,11},
Guillermo Torres\altaffilmark{5}}

\altaffiltext{1}{Astronomy Department, University of Florida, 211 Bryant Space Sciences Center, Gainesville, FL 32111, USA}
\altaffiltext{2}{SETI Institute, Mountain View, CA 94043, USA}
\altaffiltext{3}{NASA Ames Research Center, Moffett Field, CA, 94035, USA}
\altaffiltext{4}{UCO/Lick Observatory, University of California, Santa
Cruz, CA 95064, USA}
\altaffiltext{5}{Harvard-Smithsonian Center for Astrophysics, 60
Garden Street, Cambridge, MA 02138, USA}
\altaffiltext{6}{San Jose State University, San Jose, CA 95192, USA}
\altaffiltext{7}{Orbital Sciences Corp,oration/NASA Ames Research
Center, Moffett Field, CA 94035, USA}
\altaffiltext{8}{Jet Propulsion Laboratory/California Institute of
Technology, Pasadena, CA 91109, USA}
\altaffiltext{9}{Centre for Astrophysics Research, University of
Hertfordshire, College Lane, Hatfield, AL10 9AB, England}
\altaffiltext{10}{University of California, Berkeley, Berkeley, CA 94720}
\altaffiltext{11}{Bay Area Environmental Research Institute/NASA Ames
Research Center, Moffett Field, CA 94035, USA}
\altaffiltext{12}{Las Cumbres Observatory Global Telescope Network, 6740 Cortona Drive, Suite 102, Santa Barbara, CA 93117, USA}
\altaffiltext{13}{Department of Physics, Broida Hall, University of California, Santa Barbara, CA 93106, USA}
\altaffiltext{14}{Mullard Space Science Laboratory, Department of Space and Climate Physics, University College London, Holmbury St. Mary, Dorking, Surrey, RH56NT, UK}
\email{altheam@astro.ufl.edu}

\begin{abstract}

Doppler planet searches have discovered that giant planets follow orbits with a wide range of orbital eccentricities, revolutionizing theories of planet formation.  The discovery of hundreds of exoplanet candidates by NASA's {\em Kepler} mission enables astronomers to characterize the eccentricity distribution of small exoplanets. Measuring the eccentricity of individual planets is only practical in favorable cases that are amenable to complementary techniques (e.g., radial velocities, transit timing variations, occultation photometry). Yet even in the absence of individual eccentricities, it is possible to study the distribution of eccentricities based on the distribution of transit durations (relative to the maximum transit duration for a circular orbit). We analyze the transit duration distribution of {\em Kepler} planet candidates.  We find that for host stars with $T_{\mathrm{eff}} > 5100 K$ we cannot invert this to infer the eccentricity distribution at this time due to uncertainties and possible systematics in the host star densities.  With this limitation in mind, we compare the observed transit duration distribution with models to rule out extreme distributions.  If we assume a Rayleigh eccentricity distribution for {\em Kepler} planet candidates, then we find best-fits with a mean eccentricity of 0.1-0.25 for host stars with $T_{\mathrm{eff}} \le 5100 K$.  We compare the transit duration distribution for different subsets of {\em Kepler} planet candidates and discuss tentative trends with planetary radius and multiplicity.  High-precision spectroscopic follow-up observations for a large sample of host stars will be required to confirm which trends are real and which are the results of systematic errors in stellar radii.  Finally, we identify planet candidates that must be eccentric or have a significantly underestimated stellar radius.

\end{abstract}

\section{Introduction}
\label{sec_intro}
\noindent
%
The discovery of over 400 extrasolar planets with the radial velocity (RV) technique has revealed that many giant planets have large eccentricities \citep{2006ApJ...646..505B, 2008ApJ...686..621F}, in striking contrast with most of the planets in the solar system and inconsistent with prior theories of planet formation.  The realization that many giant planets have large eccentricities raises questions about whether lower-mass planets typically have nearly circular or highly eccentric orbits.

Several mechanisms have been proposed to explain the diversity of orbital eccentricities among giant planets.  For some proposed mechanisms, one would expect both terrestrial planets and giant planets to be affected similarly, in which case most rocky planets may follow eccentric orbits.  Other mechanisms are only effective at exciting eccentricities of massive planets, predicting low eccentricities for rocky planets \citep{2004MNRAS.355.1244B}.  Measuring the eccentricity distribution of rocky planets could provide empirical constraints for planet formation theories \citep[e.g.,][]{2007LNP...729..233N, 2008ApJ...686..621F} and shed light on the history of planet formation in our solar system.

A significant orbital eccentricity also affects a planet's climate (i.e., equilibrium temperature, amplitude of seasonal variability) and potentially its habitability due to variations in the incident stellar flux \citep{2002IJAsB...1...61W, 2004NewA...10...67G}.  Therefore, determining the frequency of terrestrial planets on low eccentricity orbits could influence the design of future planet searches that will attempt to characterize Earth-like planets and search for signs of life.

While discovering low-mass planets around distant stars is observationally challenging, measuring their eccentricities is even more difficult.  Most known exoplanets were discovered via the radial velocity method.  These have facilitated the characterization of the eccentricity distribution for giant planets \citep[e.g.,][]{2008ApJ...686..621F, 2008ApJ...686..603J, 2011MNRAS.410.1895Z}.  Through impressive instrument design and large observational campaigns, the radial velocity method has begun to detect ``super-Earth''-mass planets with orbital periods of less than 50 days \citep{2010Sci...330..653H, 2007A&A...469L..43U}.  In most cases, there are only weak constraints on the eccentricity, due to the limited number and precision of observations. Measurements of small eccentricities based on low amplitude radial velocity signals are further complicated by biases that favor larger eccentricities \citep{2008ApJ...685..553S, 2011MNRAS.410.1895Z, 2010ApJ...725.2166H}.  While it is likely possible to measure the eccentricity of a modest number of low-mass planets orbiting close to bright stars, we are unlikely to obtain precise measurements of eccentricities of a large number of rocky planets with the current generation of Doppler instruments and telescopes.  Similarly, merely detecting rocky planets in the habitable zone of solar-type stars is a major challenge for the Doppler technique, let alone measuring their eccentricities.  

Transit searches have also discovered over 100 exoplanets, including some smaller than 2 Earth radii \citep{2009A&A...506..287L, 2011ApJ...727...24T, Kepler11}. While transit observations present exciting opportunities to discover small planets, there are similar difficulties in measuring the eccentricities of individual planets.  Traditionally, radial velocity follow-up is required for the confirmation of transiting planet candidates.  While Doppler observations can exclude large eccentricities, measuring the eccentricity entails measuring small perturbations to the shape of the radial velocity curve. Further, most transiting planet candidates are significantly fainter than the stars typically targeted by radial velocity surveys.  Thus, the high precision and number of observations needed to measure their eccentricity would require a prohibitive amount of telescope time.

Recently, nearly continuous photometric observations by {\em Kepler} have enabled planets to be confirmed based on transit timing variations \citep[TTVs;][]{2010Sci...330...51H}.  In principle, TTVs can be quite sensitive to planet mass and orbital parameters, including eccentricities. In practice, TTV signals usually have long timescales or small amplitude, and in either case can be very complex, making it difficult to measure eccentricities from existing TTV observations \citep{2010ApJ...715..803V}.

In some cases, detailed analyses of the light curves can validate a planet candidate at a high level of confidence \citep{2011ApJ...727...24T}, even if the planet is not confirmed by an independent technique.  In other cases, small effects such as the occultation, tidal distortions, reflected light, or relativisitic beaming \citep{2003ApJ...588L.117L, 2007ApJ...670.1326Z, 2010ApJ...725L.200S} can be observed.  When observed, the time and duration of the occultation relative to the transit can provide an excellent measurement of the orbital eccentricity.  While such effects have been observed for some hot, giant planets, these signals are much more difficult to detect for small planets and/or planets at larger orbital separations.

%
Fortunately, it is possible to characterize the eccentricity distribution for a population of transiting planets, even if it is not possible to measure the eccentricity of each planet individually \citep{2008ApJ...678.1407F}.  In particular, the transit duration is inversely proportional to the orbital velocity of the planet \citep[projected onto the plane of the sky;][]{2010MNRAS.407..301K}.  If the transit duration ($D$) is longer than that expected for a planet on a circular orbit transiting across a diameter of the host star ($D_{\mathrm{circ}}$), then one can obtain a minimum eccentricity (corresponding to a transit that occurs at apocenter).  If the transit duration is shorter than $D_{\mathrm{circ}}$, then it could be due to an eccentric orbit, the planet transiting across a chord shorter than the star's diameter, or some combination of the two \citep{2005ApJ...627.1011T}.  For some high signal-to-noise transits, it is possible to measure the impact parameter ($b$), but for many small planets both $b$ and the direction of pericenter ($\omega$) are unknown.  Unfortunately, there is an approximate degeneracy between impact parameter, eccentricity and limb darkening model \citep{2009ApJ...703.1086C}, as each affects the shape of transit ingress and egress. This near-degeneracy is particularly challenging for {\em Kepler} data, since {\em Kepler's} broad bandpass leads to sizable limb darkening.  When using long-cadence data (i.e., approximately 30 minute integration time), ingress and egress profiles are further distorted.  These complications make it particularly difficult to infer the eccentricities of individual planets from {\em Kepler} data alone.  Nevertheless, the observed {\em distribution} of eccentricities can be compared to model eccentricity distributions.  When making such comparisons for a population, one must account for the effects of orbital eccentricity on transit probability \citep{2007ApJ...665L..67B, 2008ApJ...679.1566B}.

In this paper, we take the first steps towards characterizing the eccentricity distribution of small planets using photometric observations from NASA's {\em Kepler} mission \citep{2010ApJ...713L..79K}.  Our results are based on the first two ``quarters'' of {\em Kepler} data to be released February 1, 2011.  We have analyzed the light curve for each Kepler planet candidate (technically, Kepler Objects of Interest; KOIs) included in the paper accompanying this data release \citep{B11b}.  We compare the transit duration distribution for KOIs to those predicted for various eccentricity distributions and to the transit duration distributions of various subsets of KOIs.  Thus, we can begin to place limits on the eccentricity distribution of low-mass planets and investigate whether this distribution varies with planet size, orbital period, or stellar type.

As the {\em Kepler} pipeline continues to be refined and {\em Kepler} continues to collect data for over 100,000 target stars, it is likely that additional planet candidates will be identified in the coming years.  Aside from the well-understood limit on orbital periods, the incompleteness in the current sample of KOIs is expected to primarily affect signals with small signal-to-noise ratios, i.e., small planets, planets around faint stars, and planets around active stars with significant short time-scale photometric variability.  To minimize the potential effects of any incompleteness, we include an analysis of transit durations of a subset of KOIs that exceed an easily detectable signal-to-noise threshold.

The {\em Kepler} follow-up observation program (FOP) plays an important role in assessing the rate of false positives and characterizing the properties of the host stars.  An early qualitative assessment of the quality of each KOI in the form of a ``vetting flag" is provided in \citet{B11b}, and an assessment of the low false positive rate is contained in \cite{2011arXiv1101.5630M}.  To minimize the effects of potential false positives, we include an analysis of transit durations of only those KOIs currently considered likely \citep[labeled 3 in ][]{B11b}, very likely (labeled 2), or confirmed to be planets (labeled 1).

The transit duration also depends on the stellar density, which may be estimated several ways.  When possible, astroseismology can provide precise constraints.  However, this is only practical for a small fraction of Kepler target stars that are bright and/or evolved off the zero age main sequence.  

For stars somewhat less bright, high-resolution spectroscopy can measure effective temperature ($T_{\mathrm{eff}}$), metallicity ([m/H]), and surface gravity ($\log g$).  When combined with theoretical stellar models, one can infer stellar mass ($M_\star$), radius ($R_\star$), and density $(\rho_\star$), subject to potential systematic uncertainties, and with the caveat that the relation between (Teff, [m/H], log g) and a star's physical parameters is not always one-to-one, favoring a Bayesian analysis rather than a point estimate of the stellar parameters.  While the {\em Kepler} FOP has performed such observations for many target stars, many others do not have high-resolution spectroscopy at this time. Therefore, in this analysis we use an early estimate of stellar properties from the {\em Kepler} Input Catalog (KIC), based on extensive pre-launch multi-color photometry of the {\em Kepler} field \citep{Brown}.

In \S\ref{sec:datan} we describe how we have analyzed Kepler data.  We compare the distribution of durations to analytic models in \S\ref{sec:res}, paying particular attention to the duration distribution of subsets of KOIs such as small planet candidates and those in multiple planet candidate systems.  We identify a few KOIs which have extended transits potentially due to a significant eccentricity in \S\ref{sec:IndivKois}.  We discuss theoretical implications in \S\ref{sec:disc}.

\section{Methods}

We compare the normalized transit duration distribution for various subsets of {\em Kepler} planet candidates and theoretical models.  We describe our model for generating transit duration distributions in \S\ref{sec:sims}.  In \S\ref{sec:datan} we describe our reduction of {\em Kepler} data and the process of measuring the transit durations and other parameters.  We discuss the stellar parameters used for computing normalize transit durations in \S\ref{sec:kic} and our sample selection in \S\ref{sec:sample}.  

\subsection{Transit Duration Generation}
\label{sec:sims}

\begin{figure}
\includegraphics[width=\textwidth]{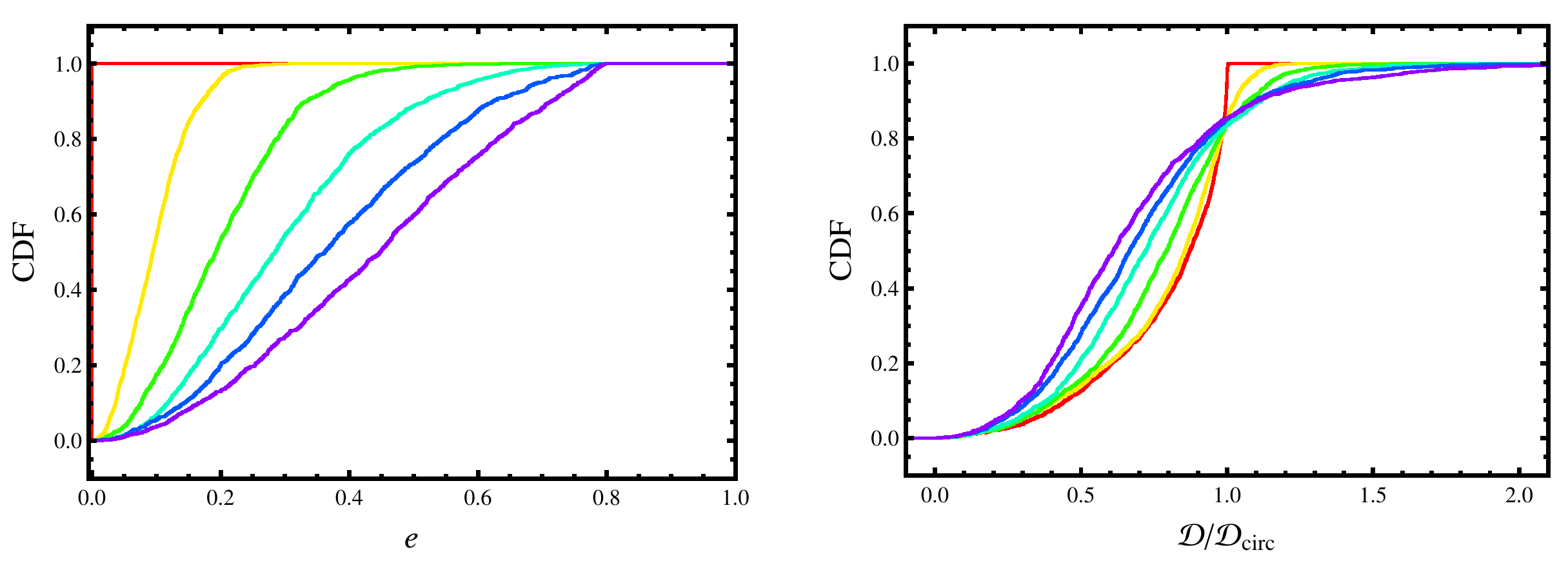}
\caption{The cumulative distribution function in transit duration (right) resulting from a Rayleigh distribution in orbital eccentricity (left).  Individual curves correspond to a varying mean for the Rayleigh distribution, ranging from $\bar{e}= 10^{-3}$ (red) to $\bar{e} = 0.5$ (violet) in increments of 0.1. }\label{roberts}
\end{figure}

We use Monte Carlo methods to generate a large sample of normalized transit durations for a given eccentricity distribution.  For our analytic models, we consider a series of truncated Rayleigh distributions for the orbital eccentricities.  The Rayleigh distribution is parameterized by either the Rayleigh parameter ($\sigma$) or the mean eccentricity ($\bar{e} = \sigma \sqrt{\pi/2}$) \citep{2008ApJ...686..603J}:
\begin{equation}
\label{rayleigh}
d N = \frac{e}{\sigma^2} \exp{\left(-{\frac{e^2}{2 \sigma^2}}\right)} d e
\end{equation}
  We consider mean eccentricities of 0.001 (essentially circular), 0.025, 0.05, ... to 0.5.  We truncate each distribution at a maximum eccentricity of 0.8. While we draw eccentricity randomly from these Rayleigh distributions, we draw the argument of pericenter from a uniform distribution and the orbital inclination from an isotropic distribution.  We reject planets if the impact parameter is greater than the stellar radius.  For those that do transit, we compute the transit duration ($D$) following \citet{2010MNRAS.407..301K}.  We define the transit to begin (end) when the center of the planet is first (last) superimposed on the limb of the star, assuming a spherical star and planet.  This definition reduces the covariance between $D$, the planet size and other transit parameters.  It also facilitates the comparison of transit durations of planets of different sizes.  We record each value of $D/D_{\mathrm{circ}}$ and construct histograms and cumulative distributions for the model eccentricity distributions (see Fig.\ \ref{roberts}) to compare with {\em Kepler} observations.

\subsection{{\em Kepler} Data Analysis}
\label{sec:datan}

Our analysis is based on {\em Kepler} photometry taken during commissioning (Q0), the first 43 days of scientific operations (Q1) and the next quarter (Q2).  For some faint stars, Q0 is not available.  The calibrated (``PA'') data contain long-timescale trends and discontinuities.  We empirically detrend the calibrated data before performing further analysis; transit data are excluded from this smoothing based on the ephemerides of \citet{B11b}.  

We exclude a time interval of width $1.5 \, D_{\mathrm{circ}}$ around the epoch of transit, where
the transit duration is given by
\begin{equation}
\label{eqnDcirc}
D_{\mathrm{circ}} = \frac{R_\ast}{\pi a} P \, ,
\end{equation}

\noindent
where $a$ is the semi-major axis, $R_\ast$ is the stellar radius, and $P$ is the orbital period.
We adopt the orbital period and stellar radius of \citet{B11b} and estimate the stellar mass obtained from the main sequence relationship of \citet{2010A&ARv..18...67T}.  Next, we smooth the remaining data over the time scale of 0.2 days or the transit duration, whichever is greater.  The data is binned into sets of this width and averaged.  We then interpolate a third order polynomial -- i.e., the minimum polynomial with which we can interpolate a smooth, continuous function -- that passes through sequential averaged data points.  Finally, all flux data -- both in and out of transit -- is normalized by this smoothing function. 

Our interpolation function is not always well-behaved near gaps in the flux data.  We treat these errors in the smoothing by discarding any affected transits in our fitting; such transits, being located near gaps in flux data, were usually incomplete to begin with.

If our initial estimate for the transit duration is incorrect (by a factor of 1.5 or more), the incorrect identification of in-transit and out-of-transit data could alter the shape of the transit.  The incorrect identification of in-transit data as out-of-transit data could cause the edges of the transit to be smeared out, which would affect our determination of impact parameter and transit duration.  The incorrect identification of large amounts of out-of-transit data as in-transit data leads to residual noise near transit edges, which inhibits the convergence of our fitting algorithm.  Therefore, we iterate the process of calculating a smoothed light curve and fitting a transit model.  Initially, we use a duration of $D_{\mathrm{circ}}$.  On the second iteration, we fold the light curve using a duration estimate based on the full-width at half depth of the normalized and folded light curve. 

Once the light curve is normalized, we phase the remaining data and fit a simplified transit model.  We assume a spherical star and planet, no luminosity from the planet, strict periodicity (unless other work has shown transit timing variations to be present), and quadratic stellar limb darkening as modeled by \citet{2002ApJ...580L.171M}.
For the sake of speed, we fit the main orbital parameters individually.  These parameters are $P$, the period, $t_0$, the time of transit, $D$, the transit duration, $p$, the planet size (in units of stellar radii), and $b$, the impact parameter, or minimum projected separation between star and planet.  We have good initial estimates for $P$ and $t_0$ from \citet{B11b}.  We obtain estimates for $D$ and $p$ using the full-width at half depth of normalized flux and the depth of the transit.  We initially assume that $b$ is zero.

Once we have these rough estimates, we loop through the orbital parameters and do an individual $\chi^2$ minimization for each.  We start with period and epoch because the fits for these quantities are largely independent of planet size and transit duration.  Next, we fit for duration and find it, too, to be fairly stable against variations in $p$ and $b$, due to our definition of transit duration.  Next, we fit for $p$, which is a proxy for transit depth.  Once $p$ is determined, we fit for $b$, varying $b$ and $p$ simultaneously so that the depth of the transit remains constant:

\begin{equation}
p(b) = p_0 \sqrt{1 - f_{p=p_0,~b=0} \over 1 - f_{p=p_0,~b}} \, ,
\end{equation}

\noindent 
where $p_0$ is the square root of the transit depth.  We iterate the fitting process to verify the stability of the solution and to refine our measurements of the orbital parameters.  For each parameter, we perform a $\chi^2$ minimization to find the best-fit parameter values and their associated uncertainties.  Note that in many cases, the uncertainty in the impact parameter is so large that we have no significant measurement.  In these cases, the degeneracy with $b$ prevents the measurement of $e \cos{\varpi}$ of an individual planet.  This motivates the use of the duration distribution as a tool to characterize the eccentricity distribution of a population of planets.

\subsection{Stellar Parameters}
\label{sec:kic}
Equation \ref{eqnDcirc} gives the maximum transit duration for a circular orbit in terms of period ($P$), semi-major axis ($a$), and stellar radius ($R_\ast$).  Transit photometry directly measures the orbital period, rather than the semi-major axis. Using Kepler's third law, we can rewrite Eq.\ \ref{eqnDcirc} using an estimate of the stellar mass ($M_\ast$) and the transit period.  Therefore, $D_{\mathrm{circ}}$ is a function of $M_\ast$, $R_\ast$, and $P$. In practice, we must estimate $D_{\mathrm{circ}}$ based on additional observations of the target star.  For nearly all KOIs considered here, the uncertainty in $D/D_{\mathrm{circ}}$ is dominated by that in $D_{\mathrm{circ}}$, which is primarily due to the unknown stellar radius ($R_{\star}$).  

\begin{figure}
\includegraphics[width=\textwidth]{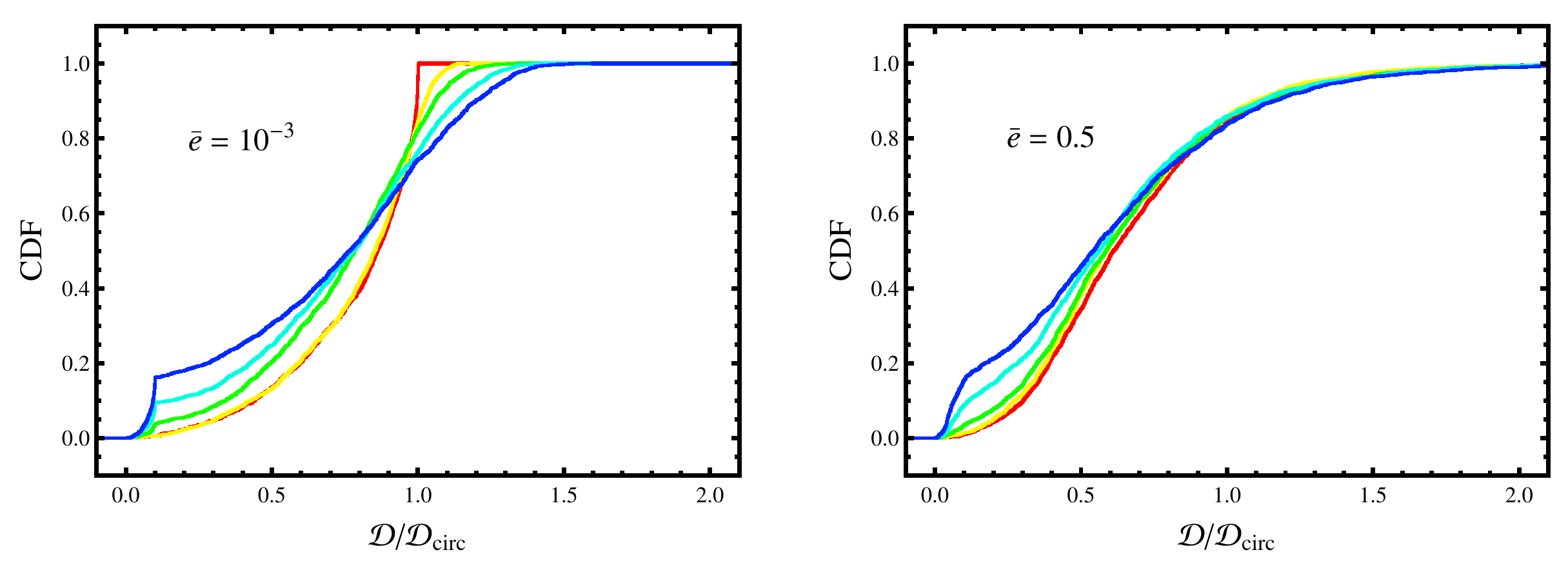}
\caption{The cumulative distribution function in noramlized transit duration resulting from a Rayleigh distribution in orbital eccentricity with a mean of $10^{-3}$ (left) and 0.5 (right), assuming zero-centered Gaussian stellar uncertainties ranging from 0\% (red) to 100\% (blue) in 25\% increments.
}\label{rsmear}
\end{figure}
\begin{figure}
\includegraphics[width=\textwidth]{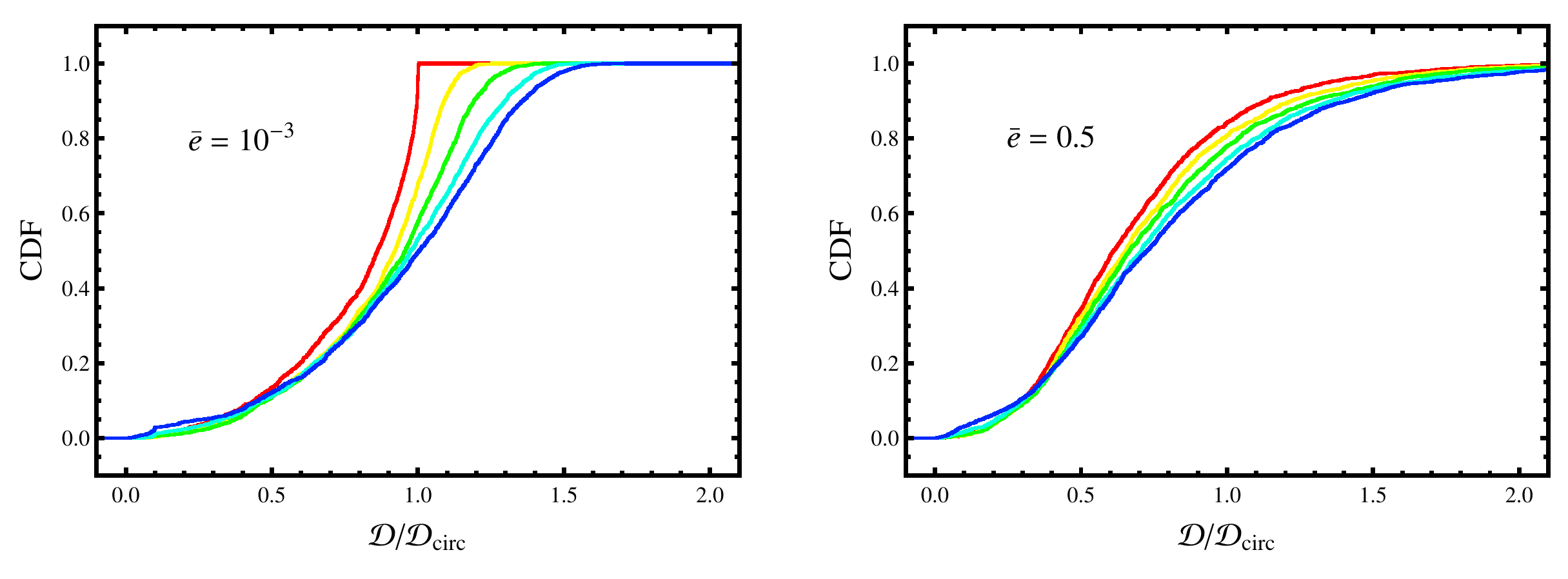}
\caption{The cumulative distribution function in transit duration resulting from a Rayleigh distribution in orbital eccentricity with a mean of $10^{-3}$ (left) and 0.5 (right), assuming Gaussian stellar uncertainties with both an offset and standard deviation ranging in magnitude from 0\% (red) to 100\% (blue) in 25\% increments.
}\label{rskew}
\end{figure}

While the Kepler Planetary Target selection function is complex, maximizing the planet detection capabilities results in favoring bright stars.  This is expected to result is a Malmquist-like bias causing more luminous subgiants to be over-represented in the Kepler planetary target list relative to a volume-limited sample.  \citet{Brown} compared KIC data and spectroscopic data for a subset of stars and recognized that the KIC often overestimates $\log(g)$ for sub-giant stars hotter than 5400 K, leading to underestimates in stellar radius by a factor of 1.5 to 2.  Since transit duration is proportional to stellar radius, an underestimated radius will result in an underestimated $D_{\mathrm{circ}}$; planet candidates will then appear to have larger $D/D_{\mathrm{circ}}$.  To investigate the sensitivity of our results to potential errors in the stellar radii, we calculate simulated transit duration distributions assuming a Rayleigh distribution of eccentricities and a Gaussian error in the stellar radius.  If the radius error is centered on zero, then the effect is to make the cumulative distribution function (CDF) shallower (Fig.\ \ref{rsmear}).  If the radius is systematically offset from the true value, then it pushes the duration CDF towards longer transit durations (Fig.\ \ref{rskew}).  Clearly, inverting the transit duration distribution to infer the eccentricity distribution will require improving the accuracy of stellar parameters.  The significant uncertainty in stellar radii and the potential for a systematic bias will drive us to focus on planet candidates associated with stars which are cooler than 5400 K, where the uncertainty in radius is reduced.

\subsection{Sample of Planet Candidates}
\label{sec:sample}

\citet{B11b} report 1217 planet candidates, each assigned a ``vetting flag'', which summarizes the level of scrutiny each planet candidate has received and the results of various tests.  The flag values correspond to: 1) a published planet, 2) a candidate that passes all tests applied, 3) a candidate that does not pass all tests cleanly, yet has no definite failures, 4) a candidate without sufficient follow-up to perform the standard vetting.  \citet{B11b} estimates the reliability of planet candidates with each vetting flag to be: 1) $\ge 98\%$, 2) $\ge 80\%$, 3) $\ge 60\%$, and 4) $\ge 60\%$.  In order to minimize contamination by potential false positives, we perform our statistical analyses only on candidates with a vetting flag of 1, 2, or 3, unless otherwise noted.   We also exclude a small number of KOIs which do not appear to have regular, symmetric transits.  For example, we exclude Kepler-9 b \& c on account of their transit timing variations (TTVs), which interfere with our systematic analysis.  This sample is expected to consist predominantly of planets which have well measured transit durations.  

As mentioned previously, we limit our analysis of transit durations of a subset of KOIs that exceed an easily detectable signal-to-noise threshold; this limit is chosen to be 15, where we have calculated the signal-to-noise ratio as the transit depth$\times \sqrt{N_{\mathrm{in\, transit}}} / $StdDev$_{\mathrm{out\, of\, transit\, flux}}$.

\section{Results} 
\label{sec:res}

In this study, we compare the distribution of normalized transit durations for KOIs to the predictions of theoretical models for the eccentricity distribution.  We also compare the normalized transit duration distribution for different subsets of KOIs to test whether there are significant differences according to properties of the planet candidates, their host stars. or quality of the data.  In each case, we use the Kolmogorov-Smirnov (K-S) test to test the null hypothesis that the two samples were drawn from the same underlying distribution.  For each test, we report the K-S statistic, $\delta$, which is defined as the maximum vertical separation between two cumulative distribution functions, and the p-value, $\alpha$, or likelihood that two samples of that size drawn from the same underlying distribution would have a K-S statistic larger than $\delta$.

\subsection{Vetting Flag}

\begin{figure}
\includegraphics[width=\textwidth]{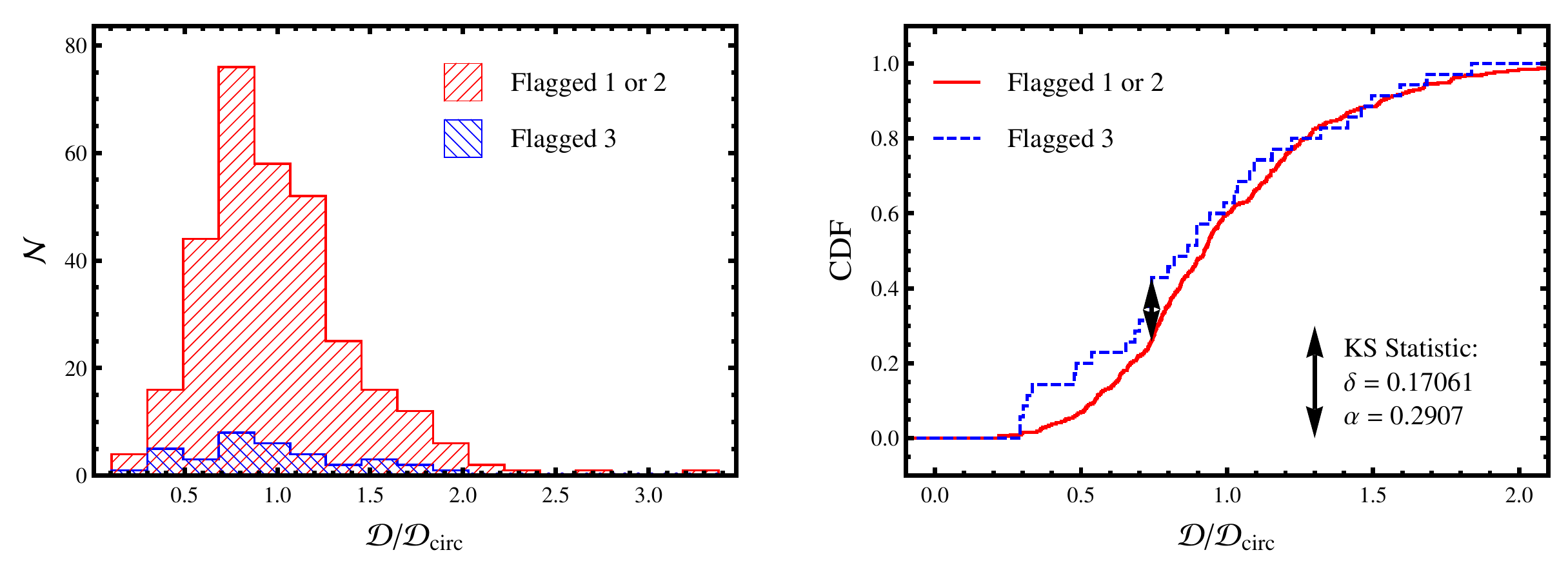}
\caption{Duration distribution (left) and cumulative distribution function (right) for planet candidates with a flag of 1 or 2 (red) and those with a flag of 3 (blue).  All suitable candidates are included, regardless of $T_{\mathrm{eff}}$.  Included in the right panel are the K-S statistic ($\delta$) and corresponding p-value ($\alpha$) between the two distributions.}
\label{flagkic}
\end{figure}

First, we compare the distribution of normalized transit durations for KOIs with different ``vetting flags'' as reported in \citet{B11b}.  As there are only 18 confirmed planets (vetting flag of 1), we compare the sample of KOIs with vetting flags of 1 or 2 ($\ge 80\%$ reliability) to KOIs with a vetting flag of 3 ($\ge 60\%$ reliability).  In Fig.\ \ref{flagkic} we compare the duration distributions of these two samples.  The K-S statistic between two samples is 0.17, and the probability that two samples drawn from the same distribution would have a statistic at least this large is $\sim 0.29$, leading us to conclude that any differences between these two sets are not statistically significant.  The two groups appear to have consistent distributions, indicating that either the false positive rate is lower than is estimated by the {\em Kepler} team or that whatever contamination by false positives is occuring does not have a discernable effect on the transit duration distribution.  Based on this result, we elect to include KOIs with vetting flags of 1, 2 or 3 in all of our subsequent analyses.

\subsection{Stellar Properties} 

\begin{figure}
\includegraphics[width=\textwidth]{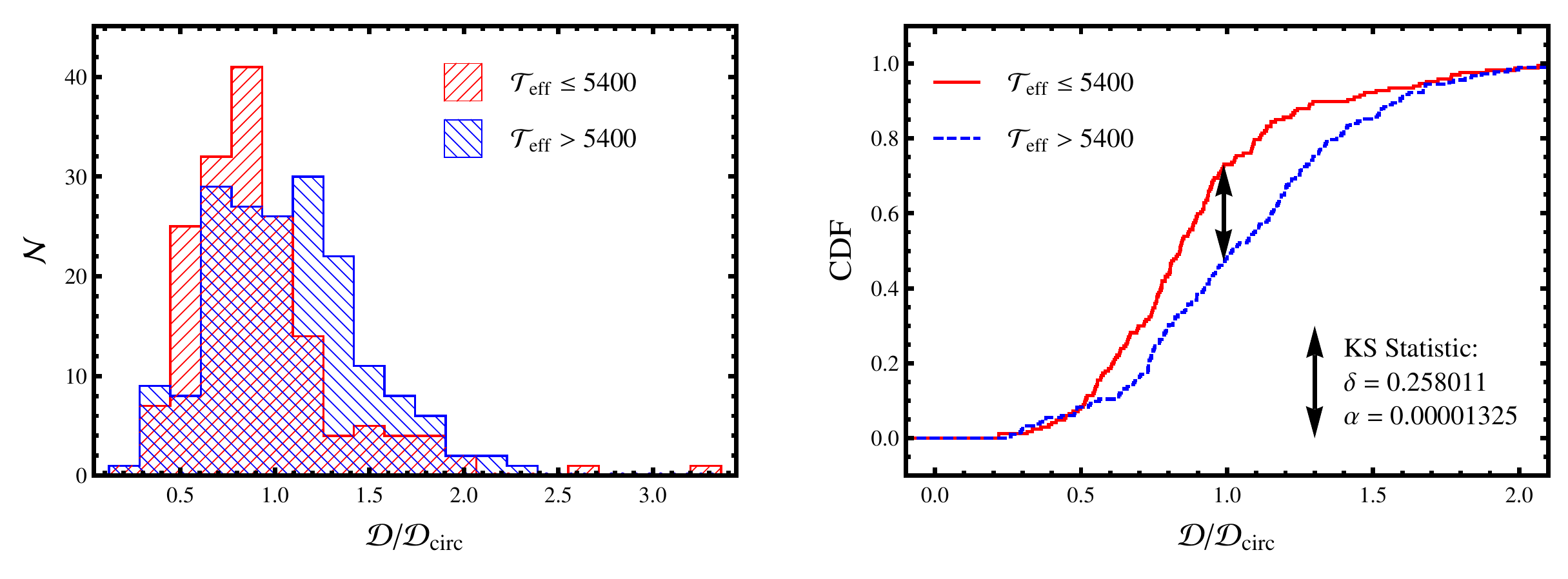}
\caption{Duration distribution (left) and cumulative distribution function (right) for planet candidates around cool (red) and hot (blue) host stars.  The dividing temperature is 5400 K.  Included in the right panel are the K-S statistic ($\delta$) and corresponding p-value ($\alpha$) between the two distributions.}
\label{teff}
\end{figure}

As discussed in \S\ref{sec:kic}, the normalized transit duration distribution can be affected by errors in the assumed stellar properties, and the largest source of uncertainty is expected to be the stellar radius.  Based on tests of a comparison sample, \citet{Brown} conclude that KIC spectroscopy cannot distinguish between hot sub-giants and main-sequence stars and hence may be significantly biased for $T_{\mathrm{eff}} > 5400$ K.  To investigate the potential for this to impact our results, we plot the distribution of normalized transit durations for two subsets of KOIs, based on whether the target $T_{\mathrm{eff}}$ is greater or less than 5400 K in Fig.\ \ref{teff}.  We find that the long duration transits (relative to $D_{\mathrm{circ}}$) are more frequent among hotter host stars.  The K-S statistic is 0.258, and the probability that the two samples were drawn from the same distribution would have a statistic at least this large is $\sim 10^{-5}$.  We conclude that there is a significant difference in the normalized transit duration between the two samples.  In principle, this could be due to a dependence of the eccentricity distribution on the stellar temperature.  However, we prefer the more cautious interpretation that the difference is more likely due to larger uncertainties in stellar radii for the hot stars.  This result is consistent with the findings of \citet{Brown} based on comparison with spectroscopic observations.  

\begin{figure}
\includegraphics[width=\textwidth]{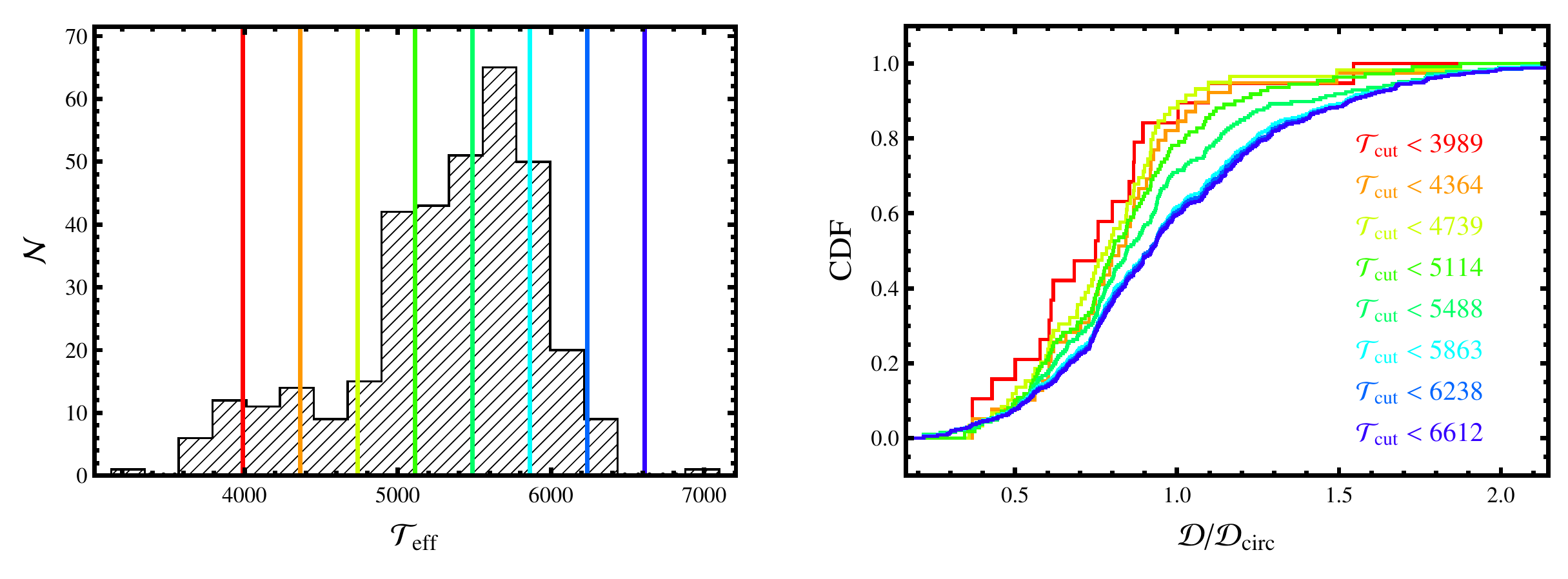}
\caption{Effective temperature histogram for host stars (left) and transit duration cumulative distribution functions corresponding to the illustrated temperature cuts for planet candidates (right).  The red CDF depicts the transit duration distribution for planet candidates around host stars for which $T_{\mathrm{eff}} < 3989$ K; each successive color includes an addition tenth of the range in $T_{\mathrm{eff}}$.}
\label{tcuts}
\end{figure}

In practice, we expect the reliability of KIC radius to change gradually as a function of temperature. Thus, we explore the duration distribution as a function of maximum effective temperature in our sample. We divide the temperature range into tenths and present CDFs for varying cutoff temperatures in Fig.\ \ref{tcuts} (although the first of these intervals does not contain enough host stars to construct a duration CDF.) We see a noticeable change in the distribution as the cutoff temperature increases from about 5000 to 6000 K and a larger number of sub-giant stars are presumably included in the sample.  As a result, we restrict our analysis to stars with $T_{\mathrm{eff}} < 5100$ K when using the duration distribution to constrain the eccentricity distribution. Unfortunately, only 20\% of KOIs satisfy this criterion, prohibiting further subdivision of the sample.  Therefore, in some cases, which are explicitly noted, we consider the normalized transit duration distribution for all stars, and when we do combine a temperature cut with candidate subdivision, we use the 5400 K value quoted by \cite{Brown}. When comparing the transit duration distribution for different subsets of planets, a modest degree of contamination is not as detrimental, since the contamination will presumably affect both samples in a similar fashion.  Of course, if we do see a difference in the duration distribution, it could be due either to a difference in the eccentricity distribution for the two subsets of planet candidates or a difference in the distribution of errors in stellar radii.

\subsection{Eccentricity Distribution of Kepler Planet Candidates}

\begin{figure}
\includegraphics[width=\textwidth]{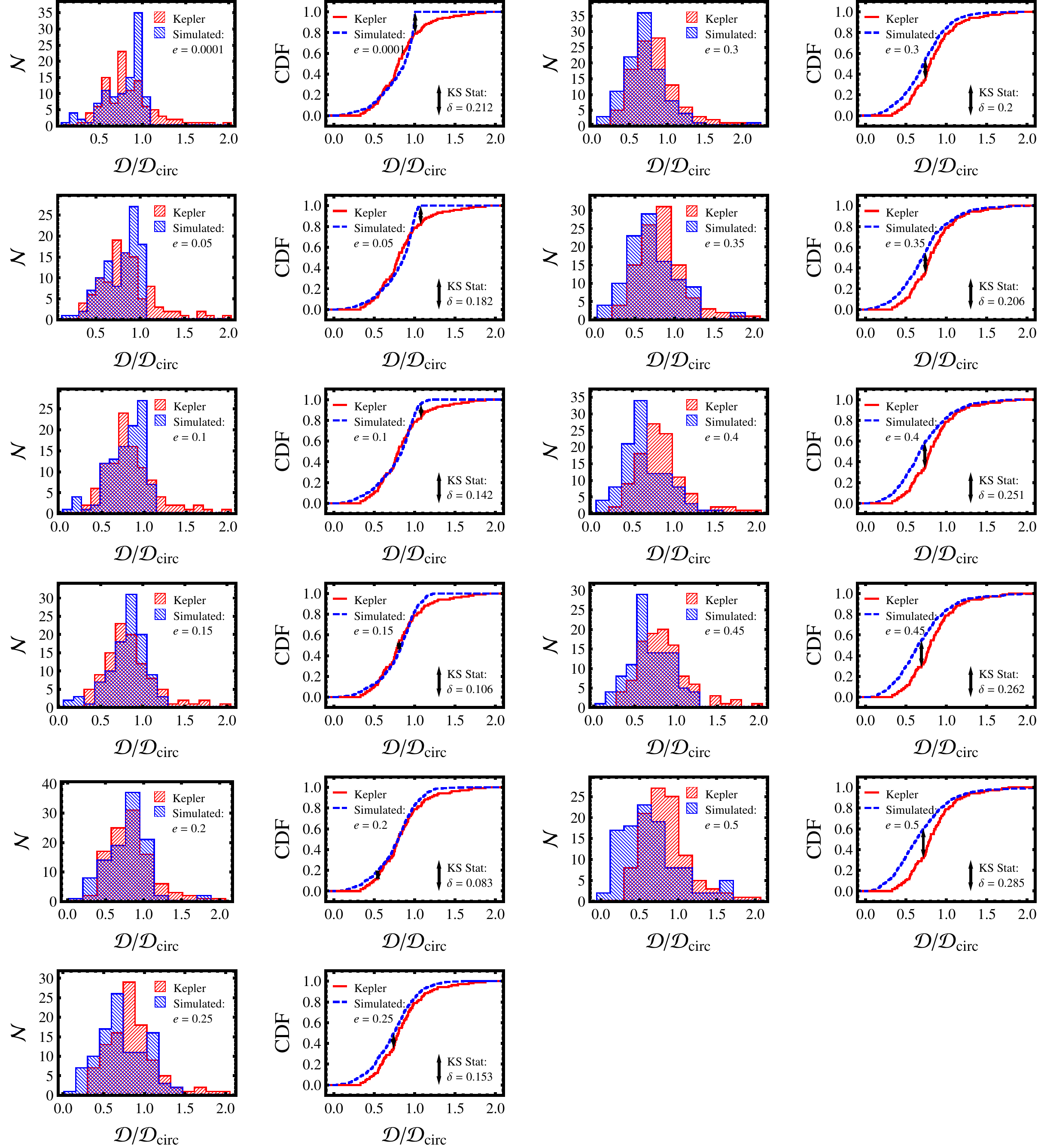}
\caption[eccarray]{Duration distributions (left) and cumulative distribution functions (right) for Kepler data (red) and simulations (blue).  Here we have restricted our data sample to those candidates around stars with $T_{\mathrm{eff}} < 5100$ K.  Each row compares the same set of Kepler data to simulated data with different underlying eccentricity distributions; each set is generated assuming a Rayleigh distribution with a different mean eccentricity, labeled on the graph.  This set shows $\bar{e} = 0.0001$ through $\bar{e} = 0.5$.  Superimposed on each CDF is the K-S statistic ($\delta$) between the two distributions.}\label{coolarray}
\end{figure}

\begin{figure}
\includegraphics[width=0.6\textwidth]{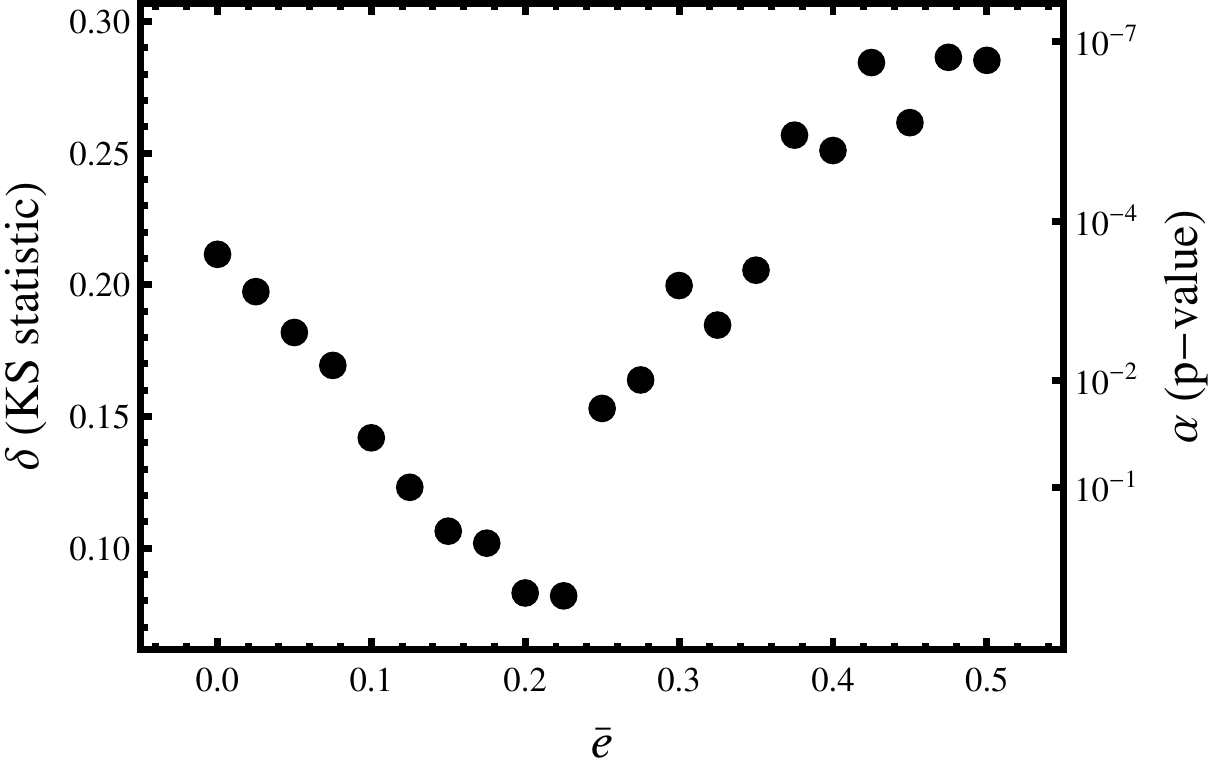}
\caption[mindeltakic]{K-S statistic resulting from the comparison of Kepler data to simulated data (Fig.\ \ref{coolarray}) as a function of the mean of the underlying (Rayleigh) eccentricity distribution of the simulated data.  The K-S statistic has a minimum of 0.08 near $\bar{e}=0.225$, corresponding to a p-value of $\alpha \sim 0.5$.}\label{coolmindelta}
\end{figure}

Based on the previous result, we investigate the normalized transit duration distribution for the subset of KOIs with host star $T_{\mathrm{eff}} \le 5100$ K.  We compare this subset to our model transit duration distributions.  Each panel of Fig.\ \ref{coolarray} shows the same histogram and cumulative distribution function for the normalized transit durations of this subsample of KOIs (red), plotted against histograms and cumulative distributions for different theoretical models (blue; see \S\ref{sec:sims}).  Each theoretical distribution corresponds to the prediction for a Rayleigh distribution of eccentricities with a mean that varies from 0.0001 (upper left) to 0.5 (bottom right).  In comparison with the small eccentricity models (e.g., $\bar{e}=10^{-4}$ or 0.05), the observed distribution has both a broader distribution and a clear overabundance of long-duration transits.  Even sizable symmetric errors in stellar radii would not resolve this discrepancy, suggesting that at least some {\em Kepler} planet candidates have significant eccentricities.  On the other hand, in comparing the observed distribution to models with a large mean eccentricity (e.g., $\bar{e}=0.4$), we find that the observed distribution has  fewer short duration events than predicted.  We plot the K-S statistic as a function of the mean eccentricity in Fig.\ \ref{coolmindelta}.  Within this family of model eccentricity distributions, we find the best agreement for a Rayleigh eccentricity distribution with a mean of $\bar{e}=0.2 - 0.225$ (with corresponding p-values of $\sim 0.5$).  If we set a minimum $\alpha$ of 0.05 for our confidence interval, our models satisfy this criterion for $0.125 \lesssim \bar{e} \lesssim 0.25$.  We reiterate, however, that this test was conducted using a relatively small subsample ($N = 104$); this analysis should be repeated as additional suitable candidates become available.

We also compare our data with the model of \cite{jipaper}, who describe the eccentricity distribution of the RV-detected planets with a two-component, Rayleigh + exponential, eccentricity distribution. The KS-statistic between this model and the data is 0.15, corresponding to a $p$-value of $\sim 0.06$, and hence also provides a possible fit to the data, though a less likely fit than our pure Rayleigh models.  Thus, the {\em Kepler} candidate transit duration distribution is not inconsistent with that of the short-period, RV-detected planets.

\subsection{Correlation with Planet Candidate Orbital Period}

Next, we investigate whether the normalized transit duration distribution provides any evidence for differences in the eccentricity distribution according to the properties of the planet candidates.  For example, theoretical models predict that giant planets with small orbital separations are likely to be circularized via tidal dissipation in the planet.  Radial velocity observations have found that most short-period ($P \le 5$d) giant planets around solar-like stars have small eccentricities (often consistent with zero), while giant planets at larger separations $P \ge 10$d often have significant eccentricities.  Since {\em Kepler} is sensitive to a broader range of planets and host stars than most previous large surveys, we consider whether the transit duration distribution could identify any differences between planets with different orbital periods or sizes.  

\begin{figure}
\includegraphics[width=\textwidth]{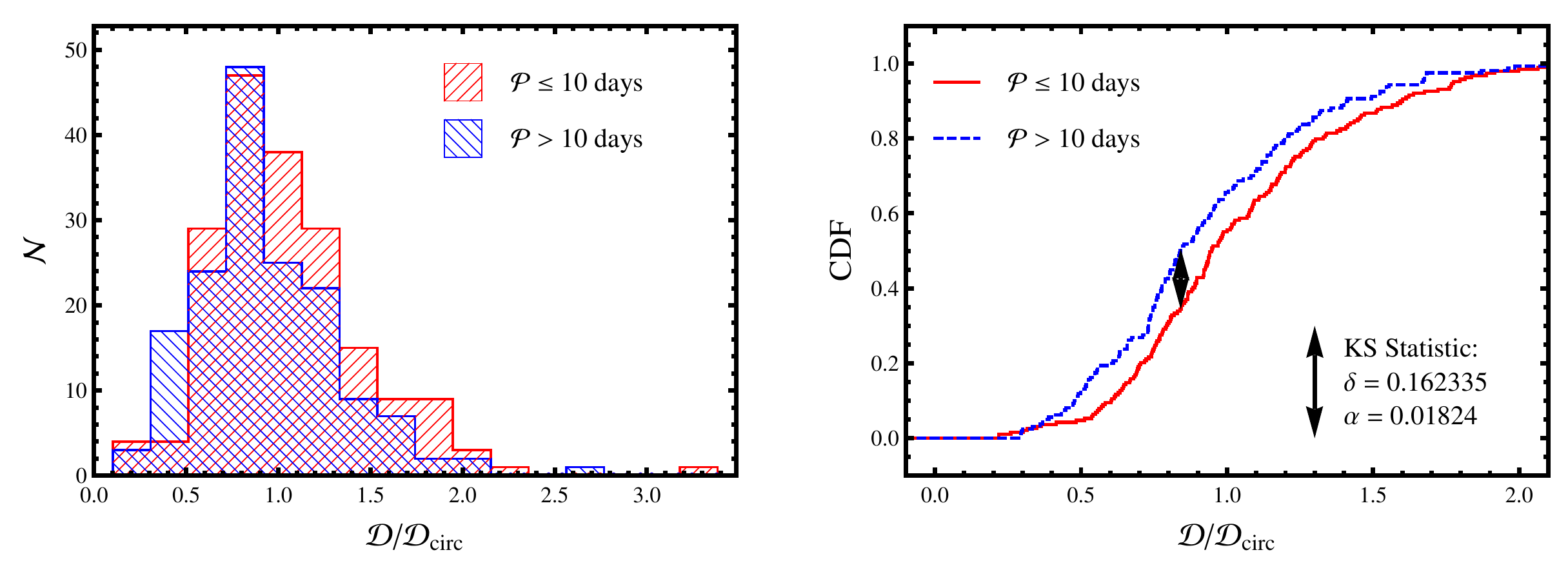}
\caption{Duration distribution (left) and cumulative distribution function (right) for planet candidates of long (red) and short (blue) orbital period.  The dividing period is 10 days, which is similar to the median period of 11 days.  No temperature cut has been applied. Included in the right panel are the K-S statistic ($\delta$) and corresponding p-value ($\alpha$) between the two distributions.}
\label{periodkic}
\end{figure}

\begin{figure}
\includegraphics[width=0.5\textwidth]{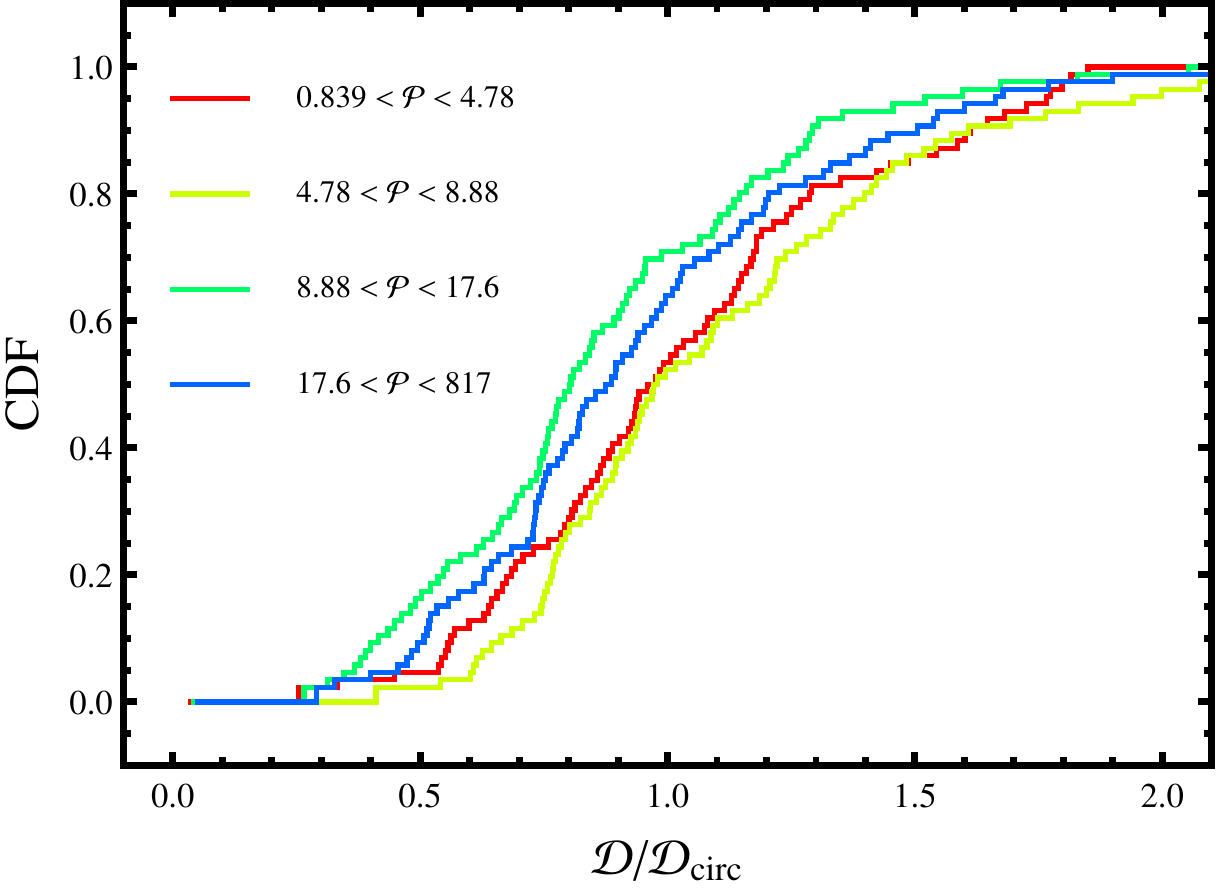}
\includegraphics[width=0.5\textwidth]{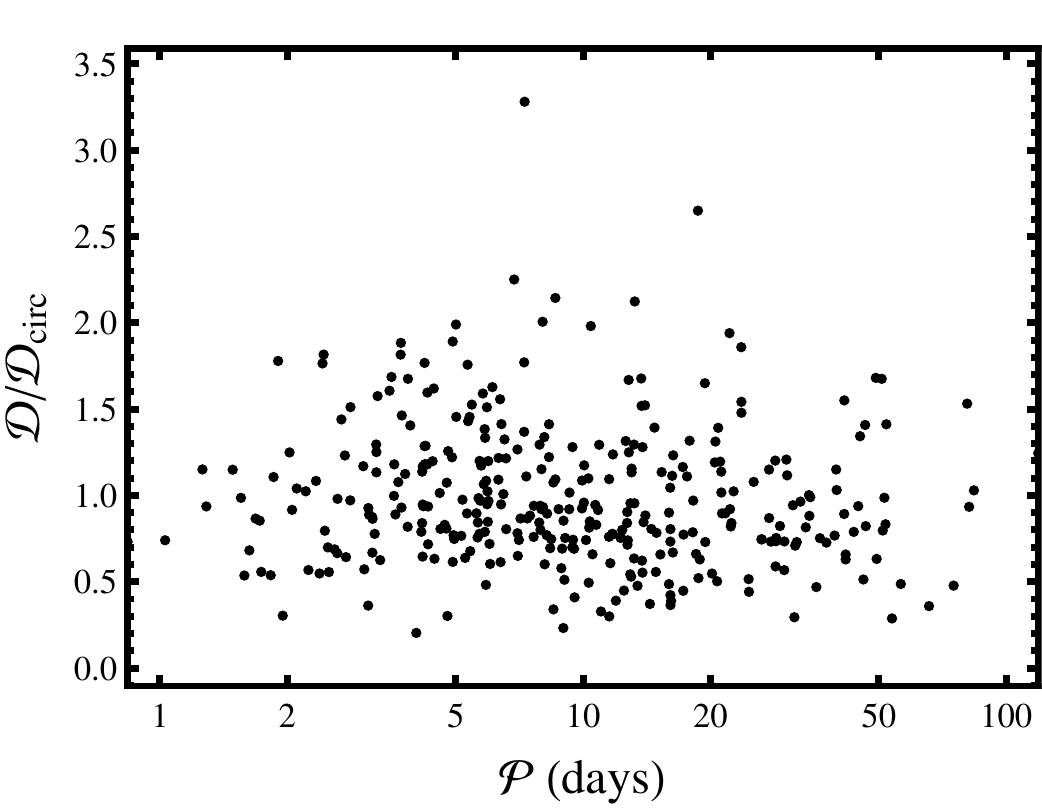}
\caption{Cumulative distribution function for planet candidates by orbital period quartile (left) and $D/D_{\mathrm{circ}}$ plotted against period for individual candidates. No temperature cut has been applied}
\label{periodqkic}
\end{figure}

First, we create subsets of KOIs based on their orbital period.  We compare the normalized transit duration distribution for KOIs with orbital period greater or less than 10d (Fig.\ \ref{periodkic}), which is close the the median 11d.  We find no significant difference, regardless of whether we restrict the samples to those with host star temperatures $\le 5400$ K.  Next, we divide the sample of KOIs into quartiles based on orbital period (Fig.\ \ref{periodqkic}).  Even the KOIs from the shortest period quartile ($P<4.78$d) do not show a significant difference from any of the other quartiles.  One would expect that this subset of candidates is more susceptible to tidal circularization, yet, if anything, this quartile tends to have a slightly higher abundance of  longer durations.  Similarly, the right panel of Fig.\ \ref{periodqkic} shows a scatter plot of $D/D_{\mathrm{circ}}$ vs.\ $P$ with no indication of a correlation between the two quantities or tidal circularization. 

\begin{figure}
\includegraphics[width=\textwidth]{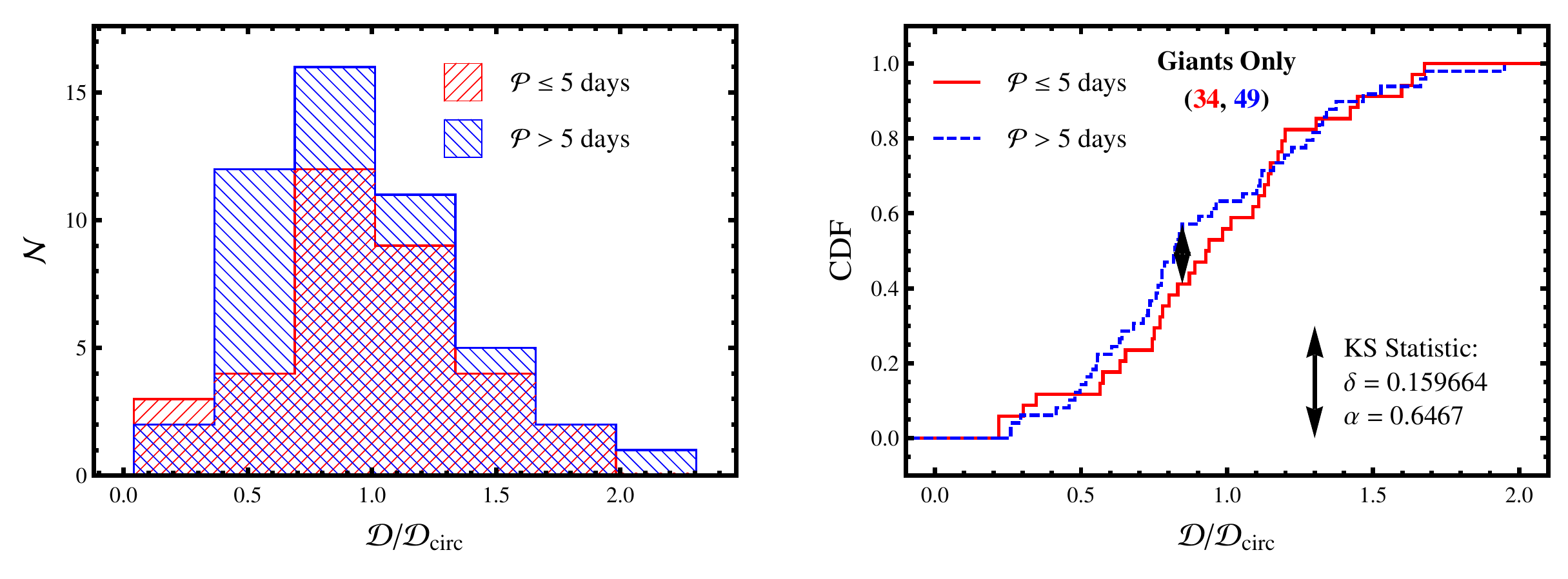}
\caption{Duration distribution (left) and cumulative distribution function (right) for giant planet candidates ($R_P > 6 R_\oplus$) of long (red) and short (blue) orbital period.  The dividing period is the tidally relevant 5 days.  No temperature cut has been applied. Included in the right panel are the K-S statistic ($\delta$) and corresponding p-value ($\alpha$) between the two distributions.}
\label{tides}
\end{figure}

One possible explanation for the lack of evidence of tidal circularization is that this phenomenon is a strong function of planet radius and most KOIs correspond to Neptune-size planets or smaller \citep{B11b}.  Therefore, we repeat our analysis for only the giant planet candidates (defined by $R_P > 6 R_{\oplus}$).  However, Fig.\ \ref{tides} shows no statistically significant difference between giant planet candidates with periods greater or less than 5 days.  (A similar comparison amongst planet candidates with $R_P > 1 R_J$ also shows no difference.)  Again, the result is not sensitive to whether we include or exclude host stars with $T_{\mathrm{eff}} > 5400$ K.  The lack of a detectable difference may be due to the relatively small number of short-period giant planets.  While {\em Kepler} has dramatically increased the number of small planet candidates, planets of Jupiter-size or greater are sufficiently rare that we have limited statistical power to discern differences in the eccentricity distribution.  Another possibility is that modest differences in the transit duration distribution are blurred out by the uncertainty in stellar radii.

\subsection{Correlation with Planet Candidate Radius}

\begin{figure}
\includegraphics[width=\textwidth]{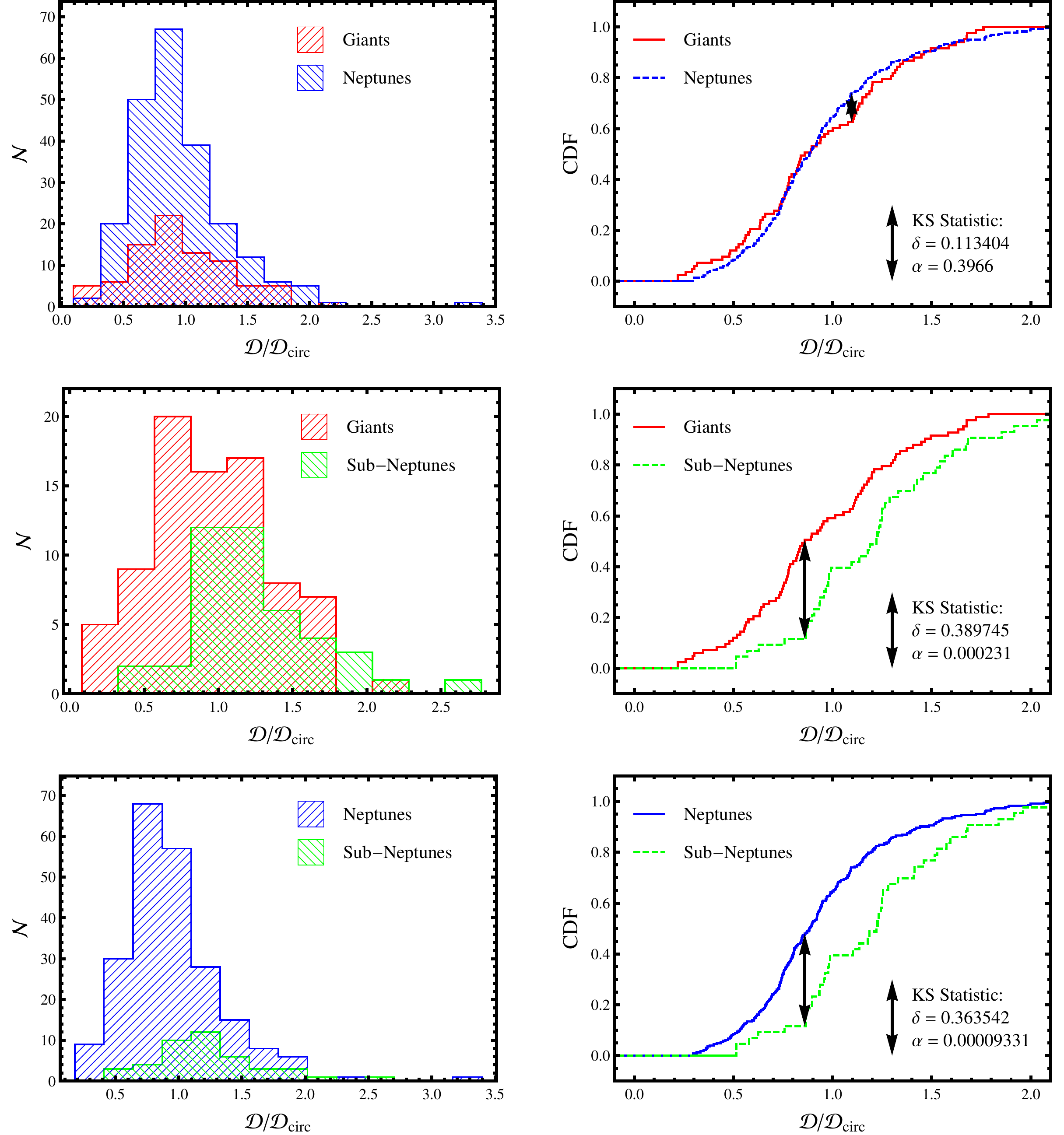}
\caption{Duration distributions (left) and cumulative distribution functions (right) for planet candidates of different size.  The three size classes shown are giants (red), Neptune-size candidates (blue), and sub-Neptune-size candidates (green).  No temperature cut has been applied. Included in the right panel are the K-S statistic ($\delta$) and corresponding p-value ($\alpha$) between the two distributions.}
\label{sizeskic}
\end{figure}

\begin{figure}
\includegraphics[width=\textwidth]{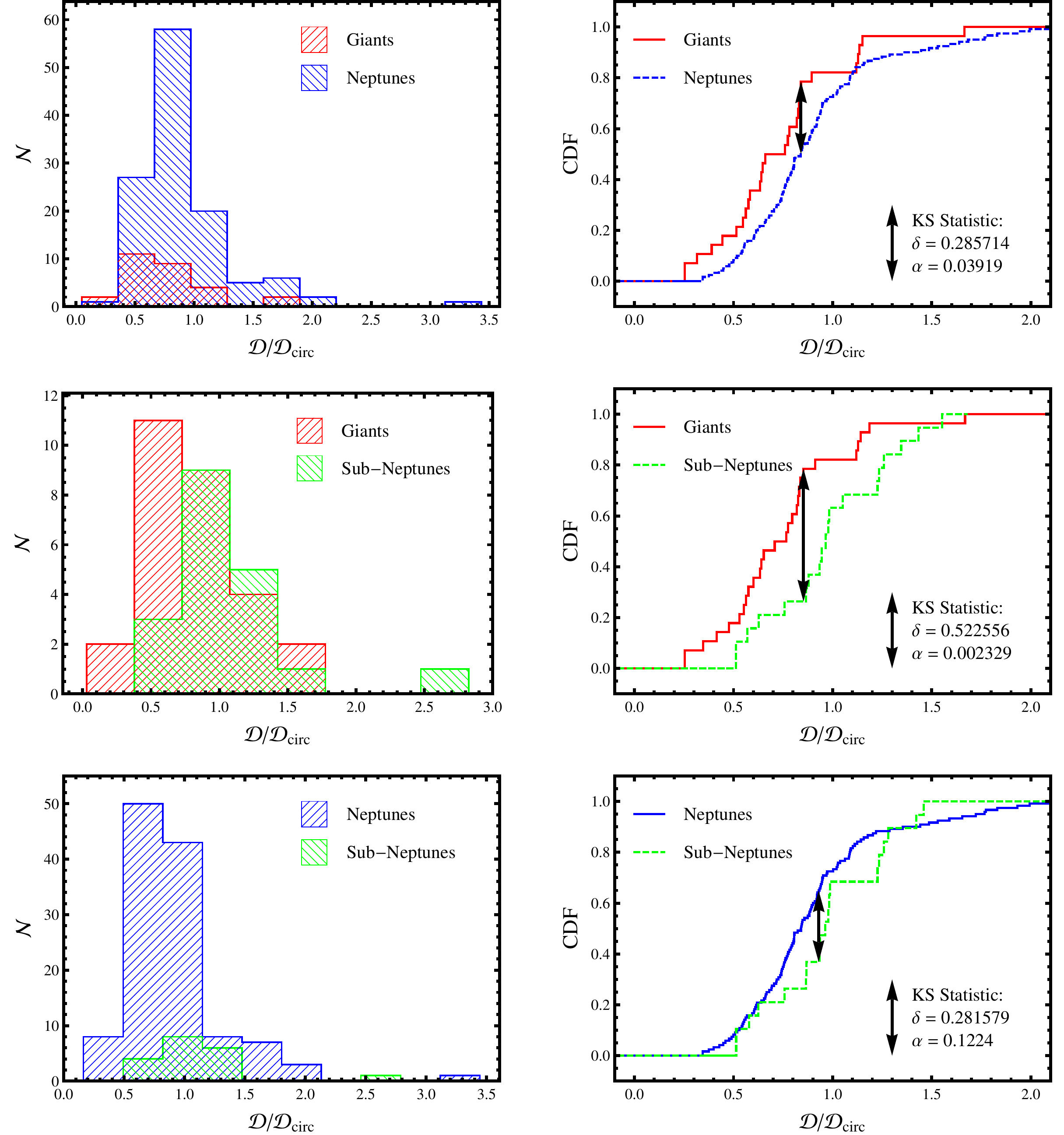}
\caption{Duration distributions (left) and cumulative distribution functions (right) for planet candidates of different size.  The three size classes shown are giants (red), Neptune-size candidates (blue), and sub-Neptune-size candidates (green).  Here we have restricted our data sample to those candidates around stars with $T_{\mathrm{eff}} \le 5400$ K.  Included in the right panel are the K-S statistic ($\delta$) and corresponding p-value ($\alpha$) between the two distributions.}
\label{coolsize}
\end{figure}

Next, we investigate whether there are differences in the normalized transit duration distribution depending on the candidate planetary radius.  Following \citet{B11b}, we separate the KOIs into three classes: giants ($R_P > 6 R_{\oplus}$), Neptune-size candidates ($2 R_{\oplus} < R_P < 6 R_{\oplus}$), and sub-Neptune-size candidates ($R_P < 2 R_{\oplus}$).  Unlike in the case of orbital period, our results for orbital size are somewhat sensitive to whether we include or exclude KOIs with hot host stars.  Therefore, we show results for both cases: no $T_{\mathrm{eff}}$ criteria (Fig.\ \ref{sizeskic}) and only KOIs with $T_{\mathrm{eff}}\le 5400$ K (Fig.\ \ref{coolsize}).  In both cases, the giant and Neptune-size candidates have statistically indistinguishable distributions.  That said, there is a hint of more frequent long duration KOIs among Neptune-size candidates in the cool stars sample.   There is a highly significant difference in the normalized transit duration distribution between Neptune-size and sub-Neptune-size planet candidates when including all stars.  However, this difference largely disappears if we consider only KOIs with host $T_{\mathrm{eff}} \le 5400$K.  The reduced statistical significance is partially due to the smaller sample size and partially due to a reduced abundance of long duration events among sub-Neptune-size planet candidates.  Regardless of which stellar sample is used, there is a statistically significant difference in the normalized transit duration distribution for giant planet candidates and sub-Neptune-size planet candidates.   Larger normalized transit durations are more common among smaller planets.  

Since the result is statistically significant even when considering only stars with $T_{\mathrm{eff}} \le 5400$ K, i.e., those which are much less prone to inaccurate radii in the KIC, it is tempting to conclude that small planets could more frequently have significant eccentricities.  However, we caution that one must consider possible systematic effects before reaching such a significant conclusion.  For example, one might worry that small planets typically have a lower SNR.  Since the SNR affects the percent uncertainty in our duration measurements, a cut in SNR is also essentially a cut in $\delta D / D$, the measurement precision for the transit duration.  Similarly, small planet candidates might be more likely to be false positives.  Yet another possibility is that small planets are more likely to be detected if they have a long transit duration.  In an effort to guard against such biases, we have only considered KOIs with a SNR greater than 15, in principle allowing the KOI to be easily detected even if its duration were shorter.  

\begin{figure}
\includegraphics[width=0.5\textwidth]{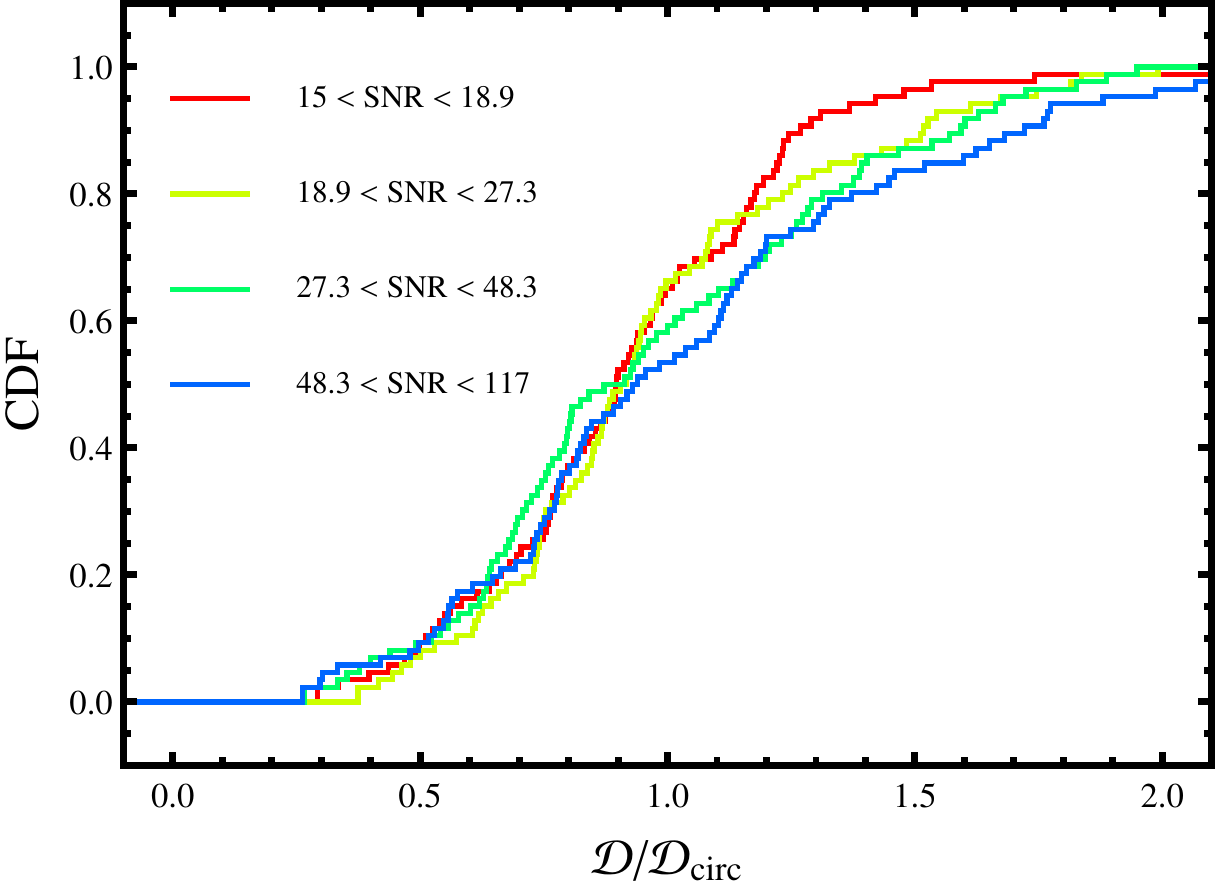}
\includegraphics[width=0.5\textwidth]{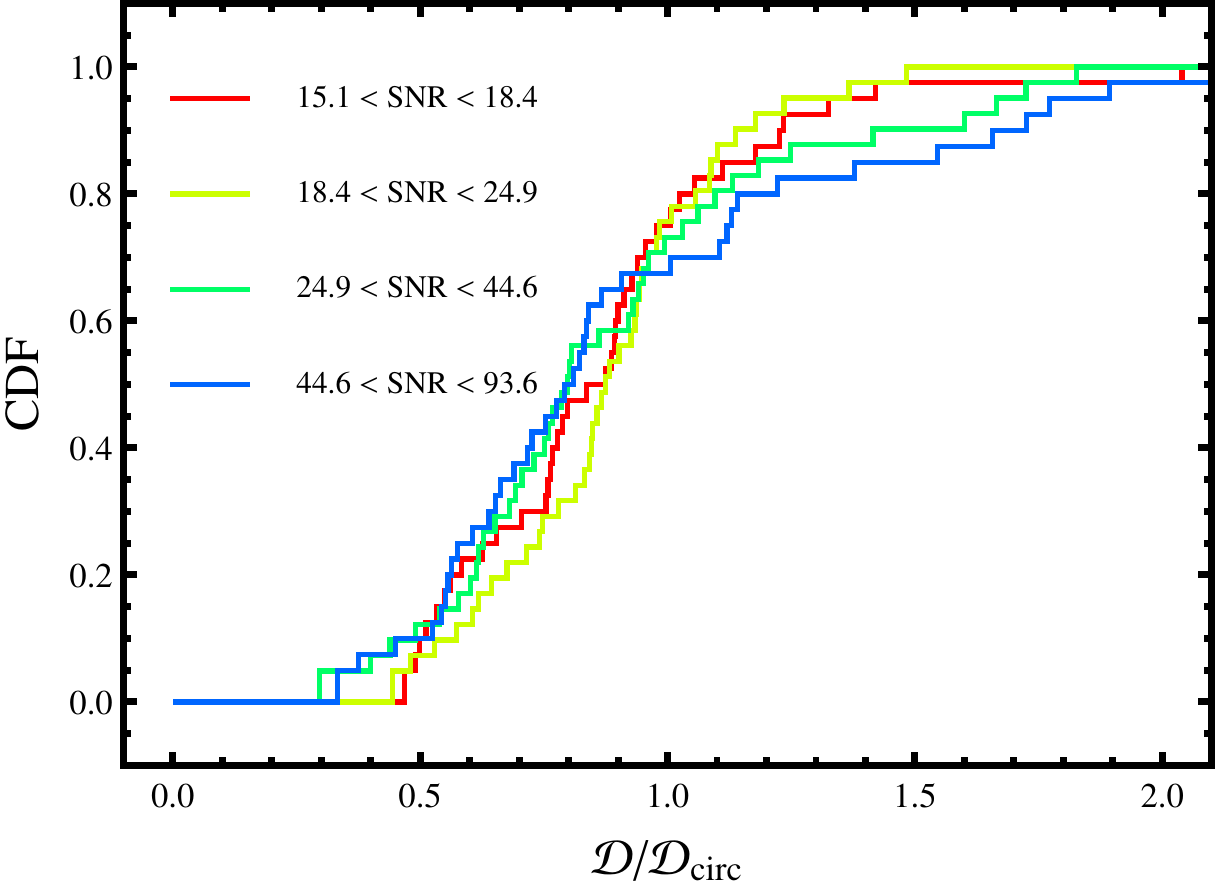}
\caption{Cumulative distribution function for {\em Kepler} planet candidates divided into quartiles by signal-to-noise ratio.  In the left panel, no temperature cut has been applied; in the right panel, we have restricted our data sample to those candidates around stars with $T_{\mathrm{eff}} \le 5400$ K.}
\label{snrq}
\end{figure}

\begin{figure}
\includegraphics[width=\textwidth]{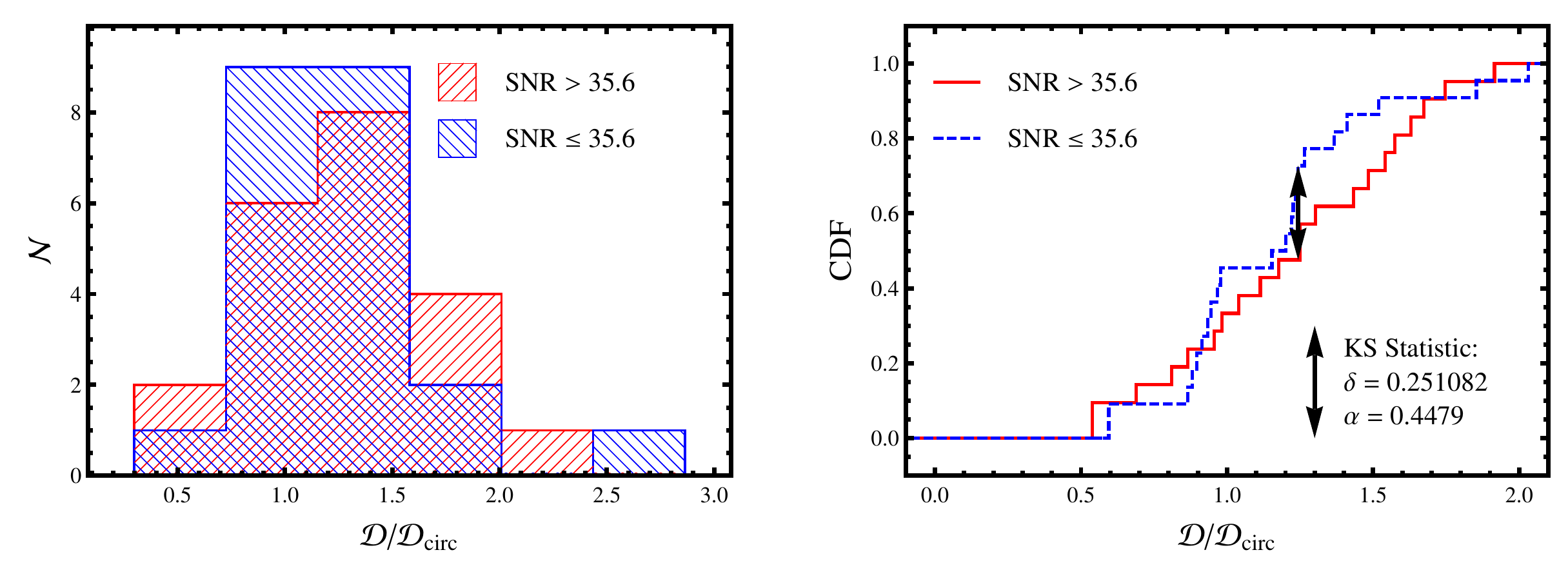}
\caption{Duration distribution (left) and cumulative distribution function (right) for sub-Neptune-size planet candidates with relatively noisy (red) and quiet (blue) lightcurves.  The dividing SNR is the median 30.1.  No temperature cut has been applied. Included in the right panel are the K-S statistic ($\delta$) and corresponding p-value ($\alpha$) between the two distributions.}
\label{snrsize}
\end{figure}

%
%
To further investigate these possibilities, we divide the KOIs into quartiles by the SNR (combined over multiple transits). Fig.\ \ref{snrq} shows the cumulative distributions for each quartile for all stars (left) and stars with $T_{\mathrm{eff}} \le 5400$ K (right).   We see the first, second and third quartiles are consistent, but we see a modest difference between these and the fourth quartile.  There is a hint that higher SNR candidates may trend towards larger normalized transit durations. These results are independent of whether we include or exclude stars with $T_{\mathrm{eff}} > 5400$ K in our analysis.  For a given stellar type, the SNR will be lowest for the smallest planet candidates.  In Fig.\ \ref{snrsize} we compare the normalized transit duration distribution for low and high SNR, focusing only on sub-Neptune-size planet candidates.  Interestingly, the sample of lower SNR KOIs (blue) conforms to the overall duration distribution while those with a higher SNR (red) appear to have a larger fraction of long-duration transits.  These results suggest that the difference in normalized transit duration distribution is unlikely due to biases or measurement errors of the planet properties.  


Another concern is whether there could be a bias due to errors in the assumed stellar properties.  When interpreting the apparent difference in normalized transit duration distribution between giant and sub-Neptune-size planet candidates, one could ask whether there could be contamination from slightly evolved hotter stars for which radii were systematically underestimated, resulting in both an underestimate of the planet size and an overestimate of the normalized transit duration.  Testing this hypothesis and identifying the root cause of the differences will require systematically obtaining high-resolution spectra of the host stars in order to improve our estimates of stellar properties for a large sample of KOI host stars.  Fortunately, the {\em Kepler} Follow-up Observation Program is working diligently to make the observations that will eventually enable such an analysis.

\subsection{Correlation with Multiplicity}

\begin{figure}
\includegraphics[width=\textwidth]{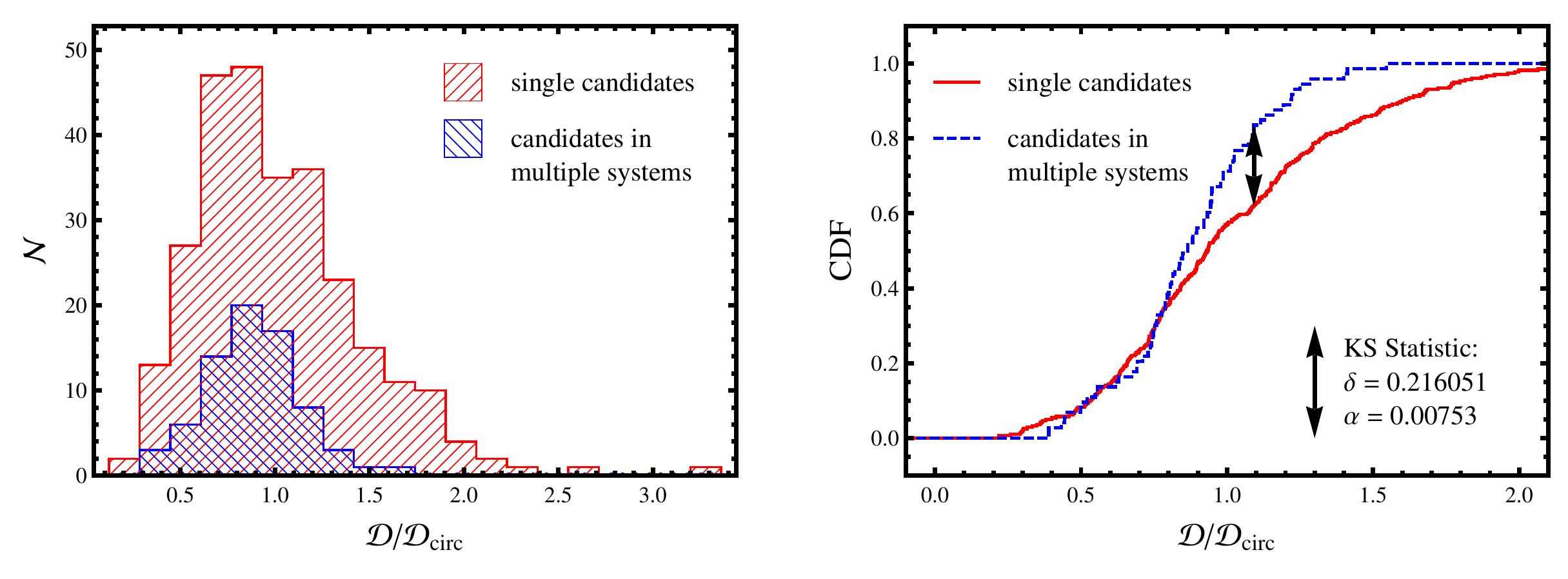}
\caption{Duration distributions (left) and cumulative distribution functions (right) for single (red) planet candidates and those in multiples (blue). No temperature cut has been applied. Included in the right panel are the K-S statistic ($\delta$) and corresponding p-value ($\alpha$) between the two distributions.}
\label{multkic}
\end{figure}

\begin{figure}
\includegraphics[width=\textwidth]{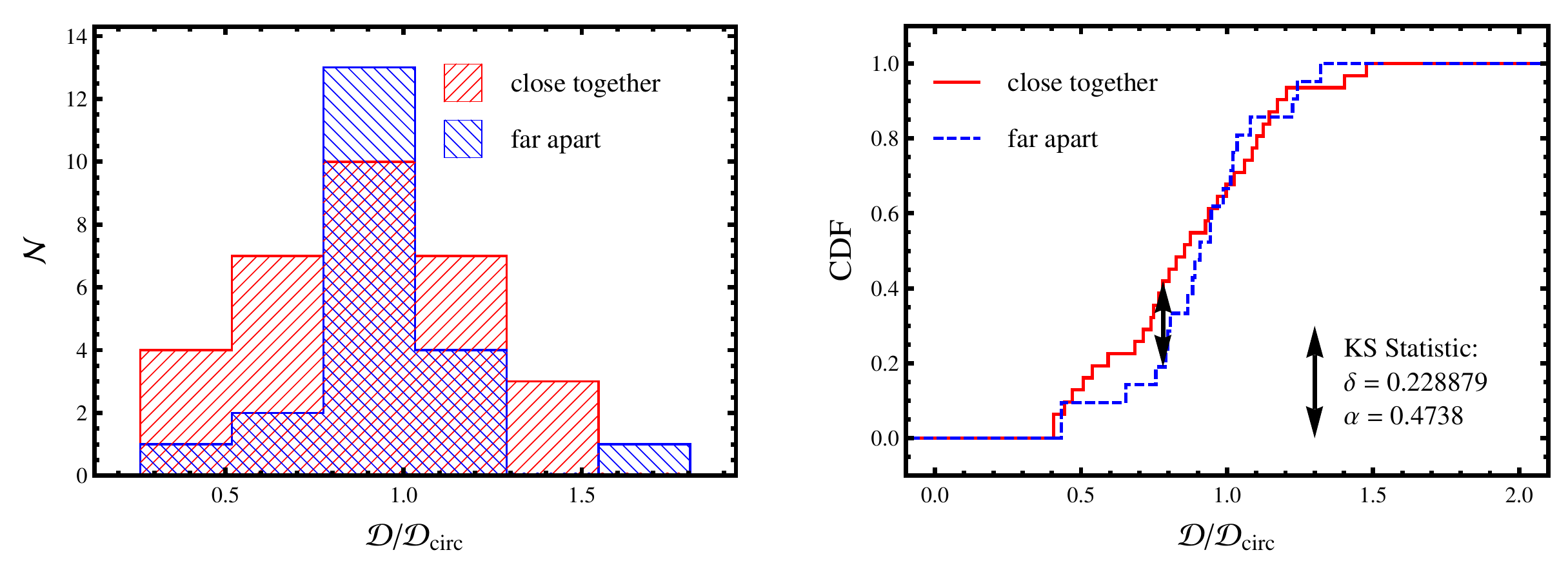}
\caption{Duration distributions (left) and cumulative distribution functions (right) for closely (red) and loosely (blue) packed planet candidates.  Only candidates in multiples are shown here, and the criterion for being considered closely packed is whether the system contains a pair of planets with a period ratio less than 2. No temperature cut has been applied. Included in the right panel are the K-S statistic ($\delta$) and corresponding p-value ($\alpha$) between the two distributions.}
\label{closevsfar}
\end{figure}

\begin{figure}
\includegraphics[width=\textwidth]{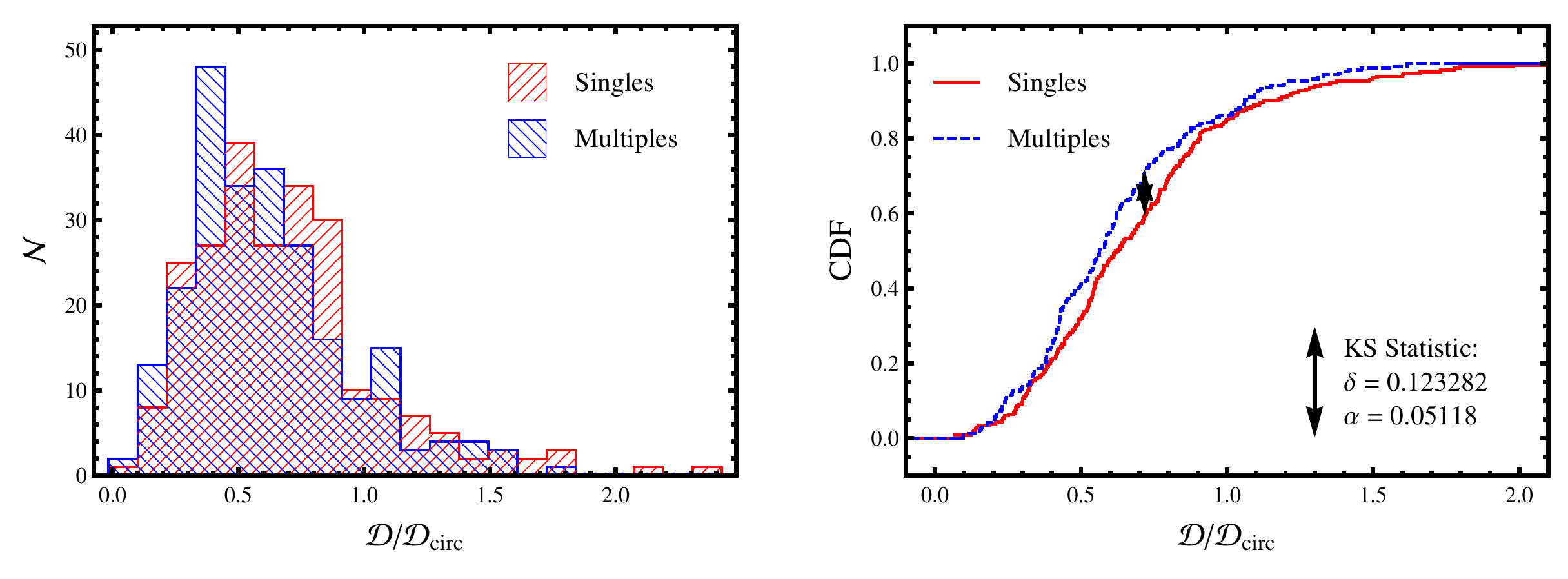}
\caption{Duration distributions (left) and cumulative distribution functions (right) for single (red) planet candidates and those in multiples (blue) in our simulated systems.  The underlying eccentricity model is a Rayleigh distribution with a mean eccentricity of 0.5.  Included in the right panel are the K-S statistic ($\delta$) and corresponding p-value ($\alpha$) between the two distributions.}
\label{noncross}
\end{figure}

Next, we investigate the effects of multiplicity, i.e., whether the distribution of transit durations differs for target stars that have multiple transiting planet candidates relative to those with just one planet candidate.  In Fig.\ \ref{multkic} we see that stars with multiple transiting planet candidates have a slightly more sharply peaked duration distribution and fewer long duration transits than that of single planet candidates.  We obtain similar results regardless of whether we include or exclude stars with $T_{\mathrm{eff}} > 5400$ K. We find a small but statistically insignificant (likely due to small sample size) difference when comparing the normalized transit duration distribution of KOIs in systems with closely or widely spaced period ratios (Fig.\  \ref{closevsfar}).  Both these results are consistent with the prediction of planet scattering models that closely-spaced multiple planet systems tend to have lower eccentricities than single planet systems.  We measure this effect in Fig.\ \ref{noncross}: we choose a model with a Rayleigh eccentricity distribution that has a mean of $\bar{e} = 0.5$, and generate planetary systems as described in \S\ref{sec:sims}.  The number of planets in each system ranges from 1 to 10.  We discard any multiple planet system which contains crossing orbits, and for each $N$ systems with $n$ planets, we keep $f N$ with $n+1$, where we have chosen $f = 0.5$.  Once these cuts have been applied, we compare this set of multiple planet systems to the single planets we generated.  We see, as expected, fewer long duration transits for the less eccentric multiples, but the effect has low significance and is small compared to the difference in {\em Kepler} candidates.  Note that we have made no stability requirements of our simulated systems; including a stability cut could increase the duration distribution's dependence on multiplicity.  

Again, we must be cautious of potential biases.  For example, multiple transiting planet systems are expected to have a lower false positive rate than KOIs with just a single transiting planet \citep{Lissauer}.  However, given the high reliability of even the single KOIs, this seems unlikely to explain the significant difference.  Another potential bias is that systems with multiple transiting planets are likely to have small relative inclinations \citep{Rag, Lissauer, Ford}.  Since most planet formation models predict that eccentricities and inclinations tend to be excited together,  systems with limited inclination excitation may also tend to have limited eccentricity excitation.  We intend to explore this possibility in a future paper, once more accurate stellar properties become available for a large sample of KOI host stars.  

\subsection{Possibly Eccentric KOIs}
\label{sec:IndivKois}

Although any of the KOIs included in our distribution analysis may be eccentric, those for which $D$ exceeds $D_{\mathrm{circ}}$ must either be eccentric or orbit incorrectly characterized host stars.  Here we identify a list of KOIs which exceed $D_{\mathrm{circ}}$ by a fair margin; all those listed below have $D - \sigma D > 1.5~D_{\mathrm{circ}}$, where $\sigma D$ is the uncertainty in $D$.  It may be desirable to obtain follow-up observations for the systems and/or obtain spectroscopic measurements for their host stars where possible.

\begin{center}
\begin{tabular}{ccccc}
KOI & $D$ & $\sigma D$ &  $D_{\mathrm{circ}}$ &  $D/D_{\mathrm{circ}}$ \\
& days & days & days & \\
 \hline
 20.01 & 0.199 & 0.004 & 0.123 & 1.619 \\
 51.01 & 0.123 & 0.013 & 0.062 & 1.981 \\
 119.01 & 0.477 & 0.046 & 0.284 & 1.681 \\
 206.01 & 0.266 & 0.025 & 0.151 & 1.757 \\
 209.01 & 0.451 & 0.028 & 0.269 & 1.675 \\
 219.01 & 0.23 & 0.029 & 0.114 & 2.006 \\
 242.01 & 0.243 & 0.036 & 0.137 & 1.771 \\
 428.01 & 0.29 & 0.036 & 0.129 & 2.251 \\
 680.01 & 0.382 & 0.023 & 0.178 & 2.144 \\
 783.01 & 0.327 & 0.072 & 0.1 & 3.281 \\
 830.01 & 0.109 & 0.004 & 0.065 & 1.686 \\
 834.01 & 0.34 & 0.054 & 0.183 & 1.859 \\
 903.01 & 0.18 & 0.028 & 0.091 & 1.99 \\
 931.01 & 0.132 & 0.008 & 0.079 & 1.675
\end{tabular}
\end{center}

Of these 14 candidates, only one is thought to be part of a multiple system (209.01); this is consistent with our earlier finding that single KOIs tend towards longer durations than those in multiples.  Of the 14, 12 candidates have host stars with $T_{\mathrm{eff}} > 5100$ K; the remaining two are 51.01, which orbits a 3240 K star, and 830.01, around a 4915 K star.

\section{Conclusions}
\label{sec:disc}

Ford et al.\ (2008) proposed using the transit duration distribution to characterize the eccentricity distribution of transiting planets. Here, we report the first such analysis for a large sample of planet candidates identified by NASA's {\em Kepler} mission.  Our results demonstrate the power of this technique for short-period planet candidates based on the first four months of {\em Kepler} data. In this first study, we use stellar radii from the {\em Kepler} Input Catalog.  We have identified several KOIs which must either have significantly eccentric orbits or inaccurate stellar radii in the KIC. Based on a comparison to spectroscopic analysis for a subsample, \citet{Brown} found the KIC to underestimate stellar radii for a significant fraction of hot sub-giant stars ($T_{\mathrm{eff}} > 5400$ K).  Our analysis supports this conclusion, as we observe a large difference in transit duration distribution between host stars above and below this threshold; there is some suggestion in our transit duration distributions indicating a lower maximum temperature may apply.

In order to minimize the effects of uncertain stellar radii, we focus our analyses on a subset of KOIs with host stars temperatures below 5100 - 5400 K.  If we assume a Rayleigh distribution for eccentricities, which is roughly consistent with the distribution of giant planet eccentricities observed by radial velocity surveys, then we find a best-fit for a mean eccentricity of $\sim 0.225$, with a p-value of 0.5.  A mean eccentricity of 0.125 to 0.25 results in a p-value above 0.05.  We caution that slight differences remain between the observations and any Rayleigh distribution, and that our determination of the eccentricity distribution is hampered by a small number of suitable candidates.  However, we can rule out extreme eccentricity distributions, such as a Rayleigh distribution with mean eccentricity greater than 0.35 or all circular orbits, if we assume that errors in stellar densities are not highly skewed for stars cooler than 5100 K.  Thus, we have promising first results indicating that the transit duration distribution can serve as an effective proxy for the underlying eccentricity distribution for a large population of planet candidates around relatively faint stars when measurements of individual eccentricities are impractical.

We compare the distribution of normalized transit durations for various subsets of the Kepler planet candidates. We find no significant difference in the transit duration distribution of KOIs assigned ``vetting flags'' of 1-2 or 3 in \citet{B11b}, suggesting that our results are robust to contamination by including some false positives along with real planets.  Similarly, we do not find any statistically significant trend with orbital period.  In particular, there is no statistically significant difference in the transit duration distribution for giant planet candidates inside of 5d.  This may be partially due to the relatively small number of such planet candidates.  While the sample size of short-period giant planets candidates is unlikely to increase significantly, we do expect that the accuracy of stellar parameters will improve as high-precision spectroscopic observations become avaliable.  Improving the accuracy of stellar densities would provide improved sensitivity to small differences in the eccentricity distribution.  Thus, we recommend that this issue be revisited when more accurate stellar parameters become available.

We also find small differences in the transit duration distribution, depending on whether there is one known transiting planet candidate or multiple transiting planet candiates for a given host star.  Improved stellar parameters for these systems will be particularly valuable, as they allow us to probe the relationship between orbital eccentricities and inclinations in multiple planet systems (Lissauer et al.\ 2011).

Finally, we find tantalizing hints of differences in the transit duration distribution as a function of planetary size.  It appears that a larger fraction of planet candidates smaller than 2 $R_\oplus$
have long transit durations than planet candidates larger than 6 $R_\oplus$.  This could be due to a larger fraction of sub-Neptune-size planets following eccentric orbits.  However, we are
particularly cautious about this result being affected by potential biases or systematic effects.  Based on comparing the transit duration distribution as a function of SNR, we do not believe it to be due to errors in the measured duration or increased probability of detecting small planets with longer durations.  In our view, the most likely bias would come from errors in the stellar parameters.  If the radii of a significant fraction of stars hosting sub-Neptune-size candidates are underestimated, then it could explain the correlation between planetary radius and transit duration.  If this is the case, then it would also imply that the planetary radii of several small planet candidates have been underestimated.  Given the considerable interest in the frequency of Earth-size planets, it is extremely important to resolve this issue.

In conclusion, we find that characterizing the eccentricity distribution of transiting planets is intimately connected to characterizing the stellar properties. The {\em Kepler} Follow-up Observation Program is making many observations to vet planet candidates and characterize their host stars.  Once these results become available for a large fraction of host stars, it will become
possible to more accurately measure host star properties, including the stellar radius.   This will increase the statistical power of tests such as those performed in this initial investigation.  For planets around bright and/or somewhat evolved stars, astroseismology can provide a precise stellar density (e.g., Batalha et al.\ 2011); this technique is particularly well-suited to bright stars, i.e., the set of stars for which we currently have the largest unresolved uncertainties, making it a very promising technique for statistical studies such as that presented here.  Targeting somewhat fainter stars with long duration transits for extended periods of short cadence observations could be used to test whether the apparently long duration is due to an underestimated stellar radius.  However, for most stars, high-precision spectroscopic observations will be needed to improve the stellar
parameters.  As {\em Kepler} finds and confirms small planets near the habitable zone, such analyses will become particularly important for understanding the frequency of Earth-size planets that could harbor liquid water on their surface.

\acknowledgements We thank Bill Cochran, Dave Latham, and Tim Brown for their many helpful suggestions and the latter two for their assistance with the {\em Kepler} Input Catalog, on which this work relies.  We thank the entire {\em Kepler} team for the many years of work that is proving so successful.  This material is based on work supported by the National Aeronautics and Space Administration under grant NNX08AR04G issued through the Kepler Participating Scientist Program.  Funding for this mission is provided by NASA's Science Mission Directorate.

\bibliography{adssample}{}

\clearpage


\begin{figure}
\includegraphics[width=\textwidth]{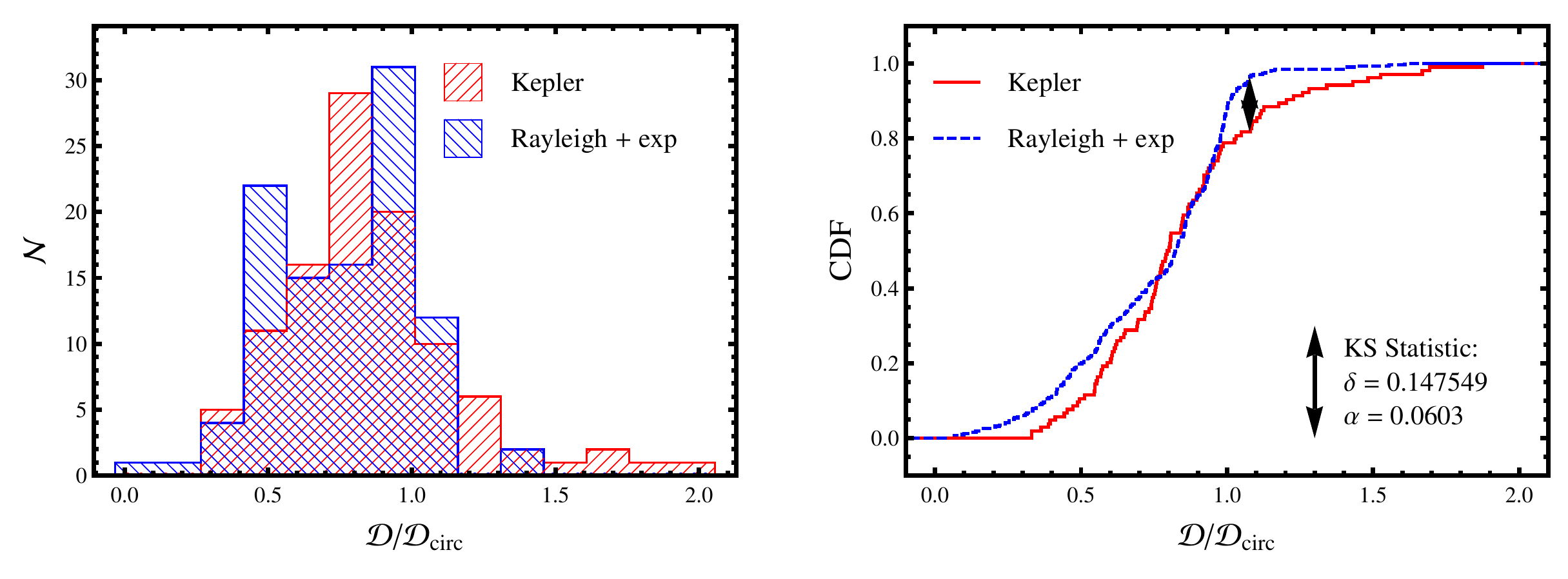}
\caption{Duration distributions (left) and cumulative distribution functions (right) for Kepler data (red) and simulations (blue).  Here we have restricted our data sample to those candidates around stars with $T_{\mathrm{eff}} \le 5400$ K.  Each row compares the same set of Kepler data to simulated data with an eccentricity sampling that follows a Rayleigh + exponential distribution, as described by \citet{jipaper}.  Included in the right panel are the K-S statistic ($\delta$) and corresponding p-value ($\alpha$) between the two distributions.}
\label{ji}
\end{figure} 

\clearpage

\newpage

\clearpage

\end{document}